\documentclass[usenatbib,fleqn,a4paper]{mnras}
\usepackage{graphicx}
\usepackage{times}
\usepackage{color}
\usepackage{amsmath}
\usepackage{amssymb}
\usepackage{geometry}
\usepackage{enumitem}

\title[Ly$\alpha$ pressure on metal-poor galaxies]
{Impact of Lyman alpha pressure on metal-poor dwarf galaxies}
\author[Taysun Kimm et al.]{  
\parbox[t]{\textwidth}{
Taysun Kimm$^{1,2}$\thanks{e-mail: tkimm@yonsei.ac.kr},
Martin Haehnelt$^{2}$,
J\'er\'emy Blaizot$^{3}$,
Harley Katz$^{2}$,
L\'eo Michel-Dansac$^{3}$,
Thibault Garel$^{3}$,
Joakim Rosdahl$^{3}$,
Romain Teyssier$^{4}$
}
\vspace*{6pt} \\
$^1$ Department of Astronomy, Yonsei University, 50 Yonsei-ro, Seodaemun-gu, Seoul 03722, Republic of Korea\\
$^2$ Kavli Institute for Cosmology and Institute of Astronomy, Madingley Road, Cambridge CB3 0HA, UK\\
$^3$ Univ Lyon, Univ Lyon1, Ens de Lyon, CNRS, Centre de Recherche Astrophysique de Lyon UMR5574, F-69230, Saint-Genis- Laval, France \\
$^4$ Institute for Computational Science, ETH Zurich, Wolfgang-Pauli-Strasse 16, CH-8093, Zurich, Switzerland 
}

\begin{document}
\maketitle

\newcommand{\fbar}{\mbox{$f_{\rm bar}$}}
\newcommand{\mbar}{\mbox{$m_{\rm bar}$}}
\newcommand{\nH}{\mbox{$n_{\rm H}$}}
\newcommand{\kms}{\mbox{${\rm km\,s^{-1}}$}}
\newcommand{\msun}{\mbox{$\rm M_\odot$}}
\newcommand{\Zsun}{\mbox{$\rm Z_\odot$}}
\newcommand{\msunyr}{\mbox{$\rm M_\odot\,{\rm yr^{-1}}$}}
\newcommand{\mvir}{\mbox{$M_{\rm vir}$}}
\newcommand{\mgas}{\mbox{$M_{\rm gas}$}}
\newcommand{\mstar}{\mbox{$M_{\rm star}$}}
\newcommand{\sfr}{\mbox{$\dot{M}_{\rm star}$}}
\newcommand{\mhalo}{\mbox{$M_{\rm halo}$}}
\newcommand{\rvir}{\mbox{$R_{\rm vir}$}}
\newcommand{\mn}{\mbox{{\sc \small Horizon}-MareNostrum}}
\newcommand{\nut}{\mbox{{\sc \small Nut}}}
\newcommand{\ramses}{\mbox{{\sc \small Ramses}}}
\newcommand{\nth}{\mbox{$n_{\rm SF}$}}
\newcommand{\cmq}{\mbox{${\rm cm^{-3}}$}}
\newcommand{\fesc}{\mbox{${\rm f_{\rm esc}}$}}
\newcommand{\Lya}{\mbox{${\rm Ly}\alpha$}}

\def\approxprop{%
  \def\p{%
    \setbox0=\vbox{\hbox{$\propto$}}%
    \ht0=0.6ex \box0 }%
  \def\s{%
    \vbox{\hbox{$\sim$}}%
  }%
  \mathrel{\raisebox{0.7ex}{%
      \mbox{$\underset{\s}{\p}$}%
    }}%
}


\newcommand{\TK}{\color{black} }
\newcommand{\mgh}{\bf \color{red} MGH: }

\begin{abstract}
Understanding the origin of strong galactic outflows and the suppression of star formation 
in dwarf galaxies is a key problem in galaxy formation.
Using a set of radiation-hydrodynamic simulations of an isolated dwarf galaxy 
embedded in a $10^{10}\,\msun$ halo, we show that the momentum transferred from 
resonantly scattered Lyman-$\alpha$ (\Lya) photons is an important source of stellar 
feedback which can shape the evolution of galaxies. We find that \Lya\ feedback 
suppresses star formation by a factor of two in metal-poor galaxies by regulating the 
dynamics of star-forming clouds before the onset of supernova explosions (SNe). 
This is possible because each \Lya\ photon resonantly scatters and imparts  
 $\sim10$--$300$ times greater momentum than in the single scattering limit.
Consequently, the number of star clusters predicted in the simulations is reduced 
by a factor of $\sim 5$, compared to the model without the early feedback. 
More importantly, we find that galactic outflows become weaker in the presence 
of strong \Lya\ radiation feedback, as star formation and associated SNe become less 
bursty. We also examine a model in which radiation field is 
arbitrarily enhanced by a factor of up to 10, and reach the same conclusion. 
The typical mass loading factors in our metal-poor dwarf system are estimated 
to be $\sim5-10$ near the mid plane, while it is reduced to $\sim1$ 
at larger radii. Finally, we find that the escape of ionizing radiation and hence the reionization 
history of the Universe is unlikely to be strongly affected by \Lya\ feedback.
\end{abstract}

\begin{keywords}
Cosmology: dark ages, reionization, first stars -- Cosmology: early Universe -- galaxies: high-redshift 
\end{keywords}

\voffset=-0.4in
\hoffset=0.0in

\section{Introduction}

Observations of local starbursts and Lyman break galaxies at high redshifts indicate 
that massive outflows are often correlated with high star formation rates 
\citep[e.g.][]{heckman15}, suggesting that stellar feedback is probably responsible 
for inefficient star formation in galaxies with $L< L_*$ 
\citep{moster13,behroozi13,sawala15,read17}. 
However, how the galactic outflows are launched and how 
star formation is regulated remain unsolved issues.

There are several different processes which can affect the dynamics of star-forming
clouds. Massive stars emit large amounts of ultraviolet (UV) photons that create 
over-pressurised regions which lower the density of the interstellar medium (ISM) 
\citep[e.g.][]{matzner02,krumholz07b}. Absorption of ionising radiation by neutral 
hydrogen and dust can also impart momentum into the ISM, driving outflows of 
$\sim 30\,\kms$ \citep{leitherer99}. Small-scale simulations of giant molecular 
clouds (GMCs) have shown that photo-heating and radiation pressure, 
caused by UV photons in the absence of dust, create a porous structure 
inside the GMCs on a timescale of a few Myr \citep{dale12,walch12,gavagnin17}.
3--40 Myr after star formation, massive stars explode as Type II supernovae (SNe), each of which release 
a large amount of energy ($\sim10^{51}\,{\rm erg}$) instantaneously. 
During the explosion, cosmic rays are generated through diffusive shock acceleration 
\citep[e.g.][]{hillas05} that may cause a pressure gradient to build up in the circumgalactic medium (CGM) 
that launches winds \citep{breitschwerdt91,everett08,hanasz13}.
In a very optically thick region, infrared (IR) photons that are re-radiated 
after the absorption of UV and optical photons can be scattered many times due to the high 
optical depths ($\tau$) and these photons impart momentum which is significantly higher 
than the single-scattering value ($L/c$) \citep{murray05,draine11b,KKO16}, where $L$ is the luminosity and $c$
is the speed of light.
Finally, radiation pressure due to resonant line radiation transfer is also known to be an important 
source of momentum to drive stellar winds \citep{castor75,abbott82}.

Of these different modes of feedback, particular attention is paid to SNe, 
as it is known to generate significantly more radial momentum compared to other sources,
such as stellar winds \citep[e.g.][]{draine11}. Local simulations of a stratified medium 
also suggest that pressure equilibrium can be maintained by injecting the momentum 
from SNe \citep{deavillez05,joung09b,kim13}, although there are uncertainties 
regarding how to model the position of SNe \citep{hennebelle14,walch15}.
On the other hand, reproducing a low star formation efficiency with SNe alone 
appears to be a much more challenging problem in a cosmological context 
\citep{joung09,aumer13,kimm15,agertz15}. This is partly because early attempts 
cannot capture the adiabatic phase of a SN explosion, during which radial momentum 
of the expanding shell is increased due to the over-pressurised medium inside 
the SN bubble. When the adiabatic phase is resolved, the final momentum is roughly 
$3\times10^5 \msun \kms$ per SN \citep{thornton98,blondin98,kim15,martizzi15}, 
which is $\sim$ 10 times greater than the momentum of initial SN ejecta. 
However, even when the correct momentum is taken into consideration, stellar 
masses of dwarf-sized galaxies still seem to be too massive \citep{hopkins14,kimm15}, 
compared to those observationally derived \citep[e.g.][]{behroozi13}.

As a solution to this over-cooling problem, \citet{murray05} put forward an idea 
that IR photons exert non-thermal pressure in the ISM and launch galactic winds.
Using isolated galaxy simulations, \citet{hopkins11} show that, when the optical 
depth ($\tau_{\rm IR}$) is estimated by the simple combination of dust column 
density and dust absorption cross-section, star-forming regions can easily attain a high $\tau_{\rm IR}$ 
of $\sim10 - 100$. If this is the case, the radiation pressure from IR photons may 
be the most important mechanism suppressing star formation and driving outflows 
in massive galaxies \citep[see also][]{aumer13,hopkins14,agertz15,agertz16}. 
However, estimating the effective optical depth that is relevant to actual radiation 
pressure does not appear to be a trivial task because dust-rich gas can be unstable 
against instabilities that create holes through which IR photons can escape 
\citep{krumholz12,davis14,rosdahl15a}. Furthermore, the inclusion of galactic 
turbulence may lead to even smaller values for optical depths \citep[e.g.][]{skinner15}.
The first galactic-scale simulations which were fully coupled with multiply scattered IR 
radiation \citep{rosdahl15b} indeed find no evidence that star formation can be controlled  
by IR pressure in halos with mass $10^{10}-10^{12}\,\msun$, even though 
$\tau_{\rm IR}$ may be under-estimated locally due to the finite resolution 
($\sim$20 pc) adopted in the study. In an opposite regime where IR radiation 
pressure is able to violently destroy the gaseous disk, the effective $\tau_{\rm IR}$ 
may again be significantly smaller than the value estimated from $\tau L/c$, 
as the photons tend to freely escape  \citep{bieri17}.

Alternatively, SNe may be able to generate stronger outflows if multiple massive 
stars explode in a short period of time. The idea behind this is to minimise the 
radiative losses in the SN shells \citep{sharma14,keller14}, although the precise 
determination of the momentum boost is still being debated 
\citep{geen16,gentry17,kim17}. Nevertheless, given that the momentum from 
SNe depends on ambient density as $\nH^{-2/17}$ it would not be surprising to 
have a more significant impact from the multiple SN events. Note that such spatially 
coherent SNe are already included in the existing simulations 
\citep{hopkins14,kimm15}. In order for super-bubble feedback to work, the extra 
momentum per SN has to be significantly larger than the momentum budget from a 
single event.  The prerequisite condition for the super-bubble feedback 
is that gas is converted into stars very quickly so that any type of early feedback does not 
intervene in the formation of massive star clusters. When a low star formation efficiency 
per free-fall time is employed in conjunction with strong IR radiation pressure, 
the ability of regulating star formation with IR photons appears to be very difficult \citep{agertz15}.
In this regard, super-bubble feedback has an implicit dependence on radiation 
feedback that operates early on, and it is desirable to fully couple hydrodynamic 
interactions to radiation in order to self-consistently model star formation and feedback.

Recently, analysing 32 HII regions in the Small and Large Magellanic clouds, 
\citet{lopez14} conclude that pressure by warm ionized gas dominates over 
UV and IR radiation pressure. The result that photoionization heating by Lyman continuum (LyC) photons
is vital for the understanding of the dynamics before the onset of SNe is also 
pointed out by several groups based on initially perturbed cloud simulations  
\citep{dale12,walch12,geen16}. The mechanism alone is unlikely to drive the strong 
outflows we observe in starburst galaxies,
but it is potentially an important source of early feedback to clear out the surrounding medium for hot bubbles 
(i.e. low-density channels) so that they can propagate easily out to the intergalactic 
medium \citep{iffrig15}.
Of course, since the Stromgren radius has a strong dependence 
on density ($r_S\propto n_{\rm H}^{-2/3}$), HII regions might be confined 
to a very small volume and would not play a role if young massive stars were deeply 
embedded in a massive GMC. 

However, both the dynamics and the role of photoionization may change 
if strong radiation pressure that has little or a positive dependence on ambient density
is present. Multiply scattered IR photons are one of those examples. 
Another interesting feedback source, which shares the similar characteristics as 
the IR pressure, is the pressure exerted by resonantly scattered Lyman 
$\alpha$ (\Lya) photons. Because $\sim68\%$ of absorbed ionizing photons 
are re-emitted as \Lya\ photons, they create a strong emission feature 
in galaxies \citep[e.g.][]{partridge67}. The main difference from IR radiation pressure is that
\Lya\ scatters mainly by neutral hydrogen, which is virtually ubiquitous in small 
galaxies regardless of metallicity. 
Moreover, since the absorption cross-section to \Lya\ is extremely high, 
the photons are unlikely to escape even when instabilities set in.
Using a simple shell model, \citet{dijkstra08} show that
a multiplication factor, the measure of momentum boost, can be very high 
($\ga100$) in optically thick regions, indicating that it may play a more important 
role than photoionization heating. Indeed, the authors demonstrate that  \Lya\ may 
be able to drive winds of several tens to hundreds of \kms\ in the ISM.
The idea is also applied to one dimensional calculations of 
dust-free halos to study the \Lya\ signature of Pop III stars \citep{smith17}. 
Despite its potential significance, little efforts have been made to understand
the impact of \Lya\ feedback on galaxy evolution. In this paper, we aim to investigate 
how \Lya\ feedback affects the evolution of star-forming clouds, the structure of 
the ISM, and whether or not it enhances galactic outflows using three 
dimensional radiation hydrodynamic simulations (RHDs) of isolated disc galaxies.

The outline of this paper is as follows.
In Section 2, we present the physical ingredients of our simulations 
and describe our \Lya\ feedback model which is based on three dimensional 
Monte Carlo radiative transfer calculations.
We analyse our simulations and quantify the impact of \Lya\ pressure 
on star formation, the properties of outflows and the ISM, and the formation 
of star clusters in Section 3. We then discuss the role of early feedback in 
launching strong galactic outflows and reionization of the Universe along with 
potential caveats and limitations of our simulations in Section 4.
Finally, we summarise and conclude in Section 5.

\section{Simulations}

We perform isolated disk galaxy simulations using the radiation hydrodynamics 
code, {\sc ramses-rt} \citep{teyssier02,rosdahl13,rosdahl15a}. 
The Euler equations are evolved using a HLLC scheme \citep{toro94}, 
with the Minmod slope limiter and a courant number of 0.7.
We adopt a multi-grid method to solve the Poisson equation \citep{guillet11}. 
For radiative transfer, we employ a first-order moment method using the M1 closure\footnote{
Although the M1 scheme is known to be more diffusive than the accurate methods, 
such as the variable Eddington tensor \citep[e.g.][]{gnedin01,davis14}, 
the expansion of HII bubble, which is a key process that determines the number 
of hydrogen recombination radiation, is well described by the M1 closure, 
as ionizing radiation from massive stars is isotropic by nature \citep{bisbas15}. 
Thus, radiative transfer with the M1 closure for the Eddington tensor is a reasonable 
choice to probe the impact of \Lya\ pressure.
}
for the Eddington tensor and the Global Lax-Friedrich intercell flux function for explicitly solving 
the advection of radiation between cells  \citep{rosdahl13,rosdahl15a}.

\subsection{Initial conditions}

Our initial condition represents a small, rotating, gas-rich, disk galaxy embedded in a 
$10^{10}\,\msun$ dark matter halo. The initial conditions were originally generated
with the {\sc makedisk} code \citep{springel05b}, and adopted previously in \citet[][G8 runs]{rosdahl15b}. The only 
difference is the gas metallicity, which is decreased to $Z=0.02\,\Zsun$ to mimic 
galaxies at high redshifts \citep[e.g.,][]{maiolino08}. The galaxy has an initially warm
($T=10^4\,{\rm K}$) gaseous disk, and is surrounded by hot ($T=10^7\,{\rm K}$), 
tenuous ($n_{\rm H}=10^{-7}\,\cmq$) halo gas. The total initial gas mass 
in the galaxy is $1.7\times10^8\,\msun$, and the initial stellar mass is 
$2.0\times10^8\,\msun$. The virial radius of the halo is $\approx$ 41 kpc, and the 
corresponding circular velocity is $\approx$ 30 \kms.

The size of the simulated box is $150\,{\rm kpc}$, which is large enough to encapsulate 
the entire halo. The computational domain is filled with $128^3$ root cells, 
which are further refined to achieve a maximum physical resolution of 4.6 pc 
($\Delta x_{\rm min}=150\,{\rm kpc}/2^{15}$). 
This is done by imposing two different refinement criteria. 
First, a cell is refined if the total baryonic plus dark matter inside each cell exceeds 
$8000\,\msun$ or if the gas mass is greater than $1000\,\msun$.
Second, we enforce that the thermal Jeans length is resolved by at least 4 cells 
until it reaches the maximum resolution \citep{truelove97}.

\subsection{Star formation and feedback}

Star formation is modelled based on a Schmidt law \citep{schmidt59},
$\rho_{\rm star} = \epsilon_{\rm ff} \rho_{\rm gas}  / t_{\rm ff}$, 
using the Poisson sampling method \citep{rasera06}, where $\epsilon_{\rm ff}$ is a star formation efficiency 
per free-fall time ($t_{\rm ff}$), and $\rho_{\rm gas}$ is the gas density.
Instead of taking a fixed  $\epsilon_{\rm ff}$, we adopt a thermo-turbulent
scheme \citep{devriendt17,kimm17} where $\epsilon_{\rm ff}$ is determined by the
 combination of the local virial parameter and turbulence, as
 \begin{equation}
\epsilon_{\rm ff} = \frac{\epsilon_{\rm ecc}}{2 \phi_t} \exp\left(\frac{3}{8}\sigma_s^2 \right) \left[ 1 + {\rm erf} \left(\frac{\sigma_s^2 - s_{\rm crit}}{\sqrt{2\sigma_s^2}}\right)\right],
\label{eq:sfe}
\end{equation}
where $\sigma_s^2 = \ln \left(1 + b^2 \mathcal{M}^2 \right)$, 
$s \equiv \ln \left(\rho/\rho_0\right)$, $\epsilon_{\rm ecc}\approx0.5$, 
$\phi_t\approx0.57$, $b\approx0.4$, and $\mathcal{M}$ is the sonic Mach number.
Here $s_{\rm crit} = \ln \left( 0.067\,\theta^{-2} \alpha_{\rm vir} \mathcal{M}^2 \right)$ approximates 
the minimum critical density above which gas can collapse, where $\alpha_{\rm vir}\equiv 2E_{\rm kin}/\left| E_{\rm grav}\right|$ is the virial parameter and $\theta=0.33$ \citep{padoan11,federrath12}.
 Note that we make a slight modification to the original scheme in order to 
allow for bursty star formation at the galactic centre. Specifically, when quantifying 
the local turbulence and virial parameter, we subtract the symmetric component of 
divergence and rotational motions. In addition, we prevent stars from forming
if a gas cell in question is not a local density maxima or the gas flow is not 
convergent. The resulting efficiency per free-fall time ranges from $\approx$ 5 to 45\% 
when a star particle is formed, with a mean $\epsilon_{\rm ff}$ of $\approx18\%$ 
in the case of the fiducial run with 4.6 pc resolution. The minimum mass of a star 
particle is $910\, \msun$, which hosts 10 individual Type II supernova explosions for a 
Kroupa initial mass function \citep{kroupa01}. SN explosions are modelled using the 
mechanical feedback scheme of \cite{kimm14} with the realistic time 
delay \citep{kimm15}. For the stellar spectra, we use the BPASS model 
\citep{stanway16} with the two different IMF slopes of -1.30 ($0.1$--$0.5\,\msun$) 
and -2.35 ($0.5$--$100\,\msun$).

\begin{table}
   \caption{Properties of eight photon groups used in the base model
   with radiation feedback excluding $\Lya$ pressure.}
   \centering
   \begin{tabular}{lcccl}
   \hline
  Photon & $\epsilon_0$ & $\epsilon_1$ & $\kappa$  & Main function \\
  group           &      [eV]          &   [eV] &   [$\rm cm^2/g$]   & \\
     \hline
   EUV$_{\rm HeII}$ & 54.42 & $\infty$ & $10^3$ & HeII ionisation\\
   EUV$_{\rm HeI}$ & 24.59 & 54.42 & $10^3$ & HeI ionisation\\
   EUV$_{\rm HI,2}$ & 15.2 & 24.59 & $10^3$ & HI and $\rm H_2$ ionisation\\
   EUV$_{\rm HI,1}$ & 13.6 & 15.2 & $10^3$ & HI ionisation\\
   LW & 11.2 & 13.6 & $10^3$ & $\rm H_2$ dissociation\\
   FUV & 5.6 & 11.2 & $10^3$ & Photoelectric heating\\
   Optical & 1.0 & 5.6 & $10^3$ & Direct RP\\
   IR & 0.1 & 1.0 & 5 & Radiation pressure (RP)\\
        \hline
   \end{tabular}
   \label{tab:photons}
\end{table}

We employ eight photon groups to account for photo-ionization heating, photo-electric 
heating on dust, direct radiation pressure from UV and optical photons, and non-thermal pressure from multi-scattered IR photons \citep{rosdahl15a}, as laid out in
Table~\ref{tab:photons}  \citep[][]{kimm17}. 
The evolution of molecular hydrogen and primordial species 
(${\rm HI}$, ${\rm HII}$, ${\rm HeI}$, ${\rm HeII}$, ${\rm HeIII}$, ${\rm e^{-1}}$, and ${\rm H_2}$)
is followed by solving 
the non-equilibrium photo-chemistry (see \citealt{katz17} and \citealt{kimm17} for details). 
To keep the computational costs low, we use a reduced speed of light approximation
 ($\tilde{c} = 10^{-3} c$), where $c$ is the full speed of light.
A uniform ultraviolet background \citep{haardt12} is included with the 
self-shielding approximation \citep{kimm17} assuming that the simulated galaxies are at $z=3$.

\begin{figure*}
   \centering
   \includegraphics[width=7.8cm]{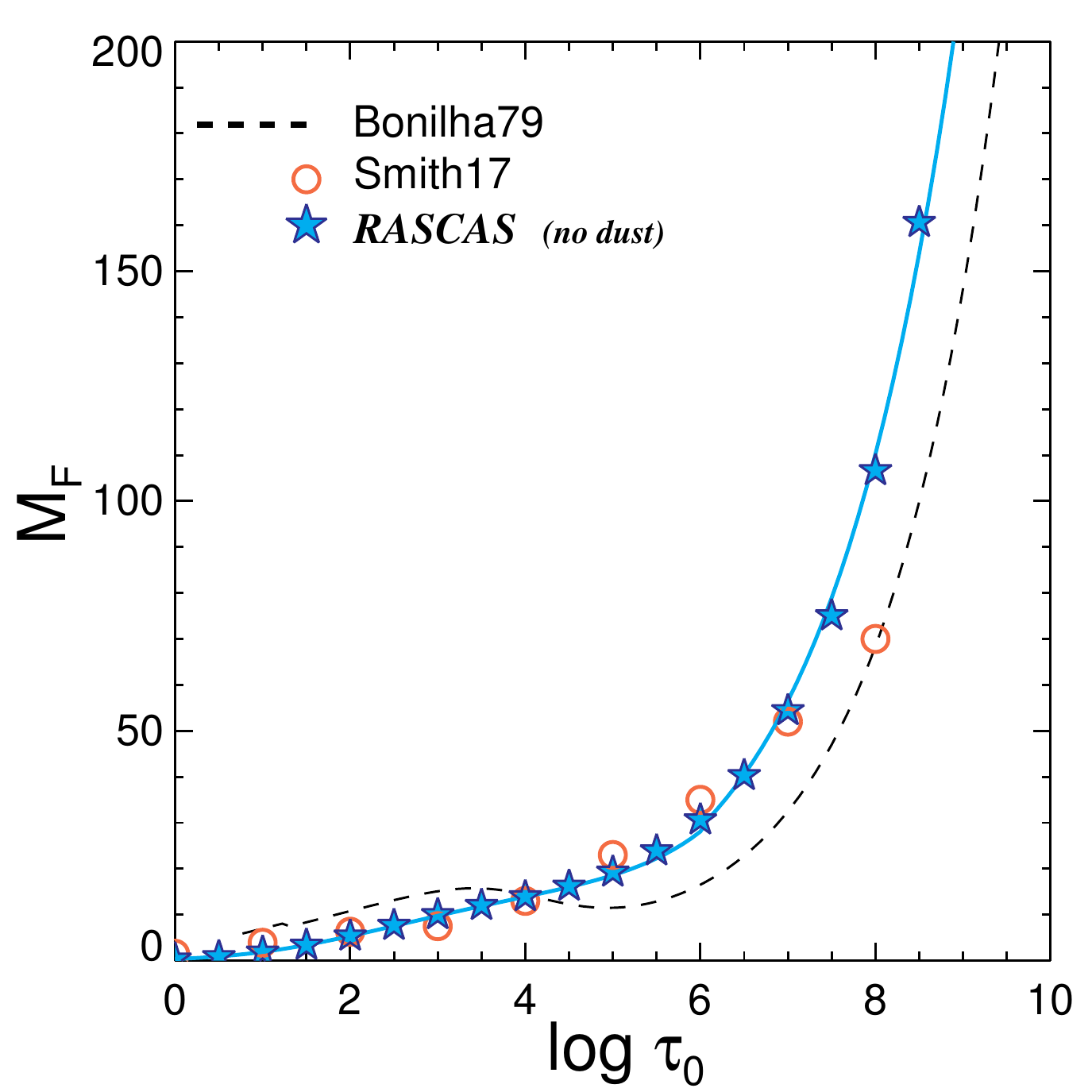} 
   \includegraphics[width=7.8cm]{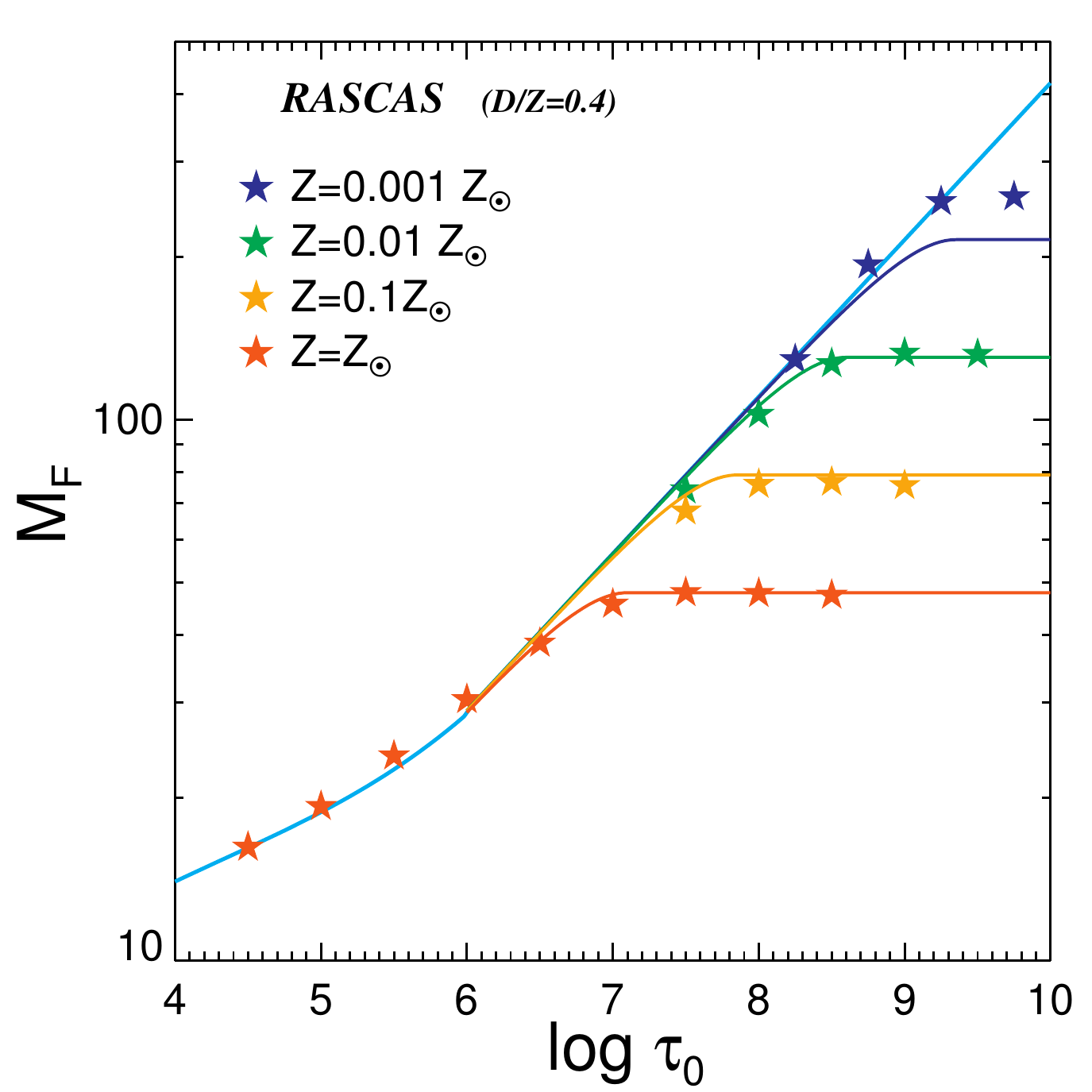} 
   \caption{ Multiplication factor ($M_{\rm F}$) of \Lya\ photons as a function of 
   optical depth at the line centre ($\tau_0$), computed using the Monte-Carlo \Lya\ 
   radiative transfer code, {\sc rascas}. {\it Left}: $M_{\rm F}$ in the dust-free medium 
   with temperature $T=100\,{\rm K}$. The dashed line indicates the results ($t_{\rm trap}/t_{\rm light}\approx M_F$) from an 
   earlier work by \citet{bonilha79} with temperature $T=10^4K$, and the empty circles 
   correspond to the results ($t_{\rm trap}/t_{\rm light}$) at $T=1K$ from \citet{smith17}. The solid line indicates 
   the fit to our calculations (Equations~\ref{eq:mf1}-\ref{eq:mf2}). 
   {\it Right}: $M_{\rm F}$ in the presence of dust.  The star symbols with different 
   colour-codings are the results from our Monte Carlo calculations with different 
   metallicities assuming a constant dust-to-metal ratio of 0.4. The solid lines are 
   the fits to these results (see the text). Note that each \Lya\ photon can transfer 
   $\sim$10--300 times larger momentum than the single scattering case in metal-poor
   environments.
   }
   \label{fig:mfactor}
\end{figure*}

\subsection{Lyman-$\alpha$ pressure}

\subsubsection{Model}
 \Lya\ photons are known to resonantly scatter in optically thick regions due to 
 the large absorption cross-section of neutral hydrogen before they escape or get absorbed by dust 
  \citep{osterbrock62,adams72,bonilha79,neufeld90}.
The number of scatterings in a dust-free medium are typically on the order of $\tau_{0}$, 
the optical depth from the centre to the edge at the line centre 
\citep[e.g.][]{adams72,harrington73,dijkstra06}. This implies that 
the momentum transfer from the scattering by \Lya\ photon can potentially be 
very important in the evolution of star-forming clouds.

The momentum transfer from resonant line transitions is usually expressed 
in terms of a multiplication factor ($M_F$), as
\begin{equation}
F_{Ly\alpha} = M_{\rm F} \frac{L_{Ly\alpha}}{c},
\end{equation}
where $F_{Ly\alpha}$ is the force due to \Lya\ photons of luminosity $L_{Ly\alpha}$.
Note that the optical depth to \Lya\ is very sensitive to the exact frequency of a photon,
and, therefore, is not simply proportional to the column density of a scatterer, 
as is often done for IR photons.
The multiplication factor can be thought of as the ratio between the time $\Lya$ photons are trapped 
and the light crossing time ($t_{\rm light}$). 
Early works showed that $M_F$ increases rapidly with 
increasing optical depth in a dust-free slab of neutral hydrogen  \citep{adams72,bonilha79}, 
resulting in  \citep[see][]{dijkstra08}, 
\begin{align}
M_{\rm F}^{\rm nodust} \approx &~ t_{\rm trap} / t_{\rm light} \\
    \approx &\,15 \times \left(\tau_{0} /10^{6} \right)^{1/3} \, 	T_4^{1/6}~~~(\log \tau_{0} \ge 6),
\end{align}
where $t_{\rm trap}$ is the trapping time, $t_{\rm light}$ is the light crossing time of the slab,
and $T_4\equiv T/10^4\, {\rm K}$. 
The dependence of ${1/3}$ on the optical depth may be understood as a consequence of coherent scattering 
in an extremely optically thick regime \citep{adams75}.

In order to probe a wider parameter space and the effects of dust, 
we measure $M_{\rm F}$ using a new three-dimensional Monte Carlo \Lya\ radiative transfer code, 
{\sc rascas} (Michel-Dansac et al. {\sl in preparation}), which is based on \citet{verhamme06}.
Monte Carlo radiative transfer (MCRT) calculations are performed with the recoil effects, 
the dipolar angular redistribution functions, and scattering by deuterium with the 
abundance ($D/H=3\times10^{-5}$). We adopt a dust albedo of $\mathcal{A}_b=0.46$ 
and the extinction cross-section per hydrogen nucleus of 
$\sigma_d = 3\times10^{-21}\,{\rm cm^2/H}$ at solar metallicity ($Z_\odot=0.02$).
The latter is slightly higher than \citet{weingartner01} and is thus a conservative estimate 
in terms of \Lya\ pressure. 

We first place
a \Lya\ emitting stellar source at the centre of a uniform medium 
with temperature $T=100\, {\rm K}$ and compute the momentum transfer at each 
scattering event from $N_{\rm Lya}=10^4$ photons, as
\begin{equation}
\Delta \vec{p} =\frac{h_p}{c} \left( \nu_{\rm in} \hat{n}_{\rm in} - \nu_{\rm out} \hat{n}_{out} \right).
\end{equation}
Here $\nu_{\rm in}$ and $\nu_{\rm out}$ are the frequency before and after 
a scattering, $\hat{n}$ denotes a direction vector, and $h_p$ is the Planck constant.
We then measure the {\em net radial momentum} centred on the stellar source, 
and divide it by the momentum due to a single \Lya\ scattering event to obtain $M_{\rm F}$.
The resulting $M_{\rm F}$ is shown as a function of 
optical depth ($\tau_0=\sigma_0\, N_{\rm HI}$) at the line centre 
($\nu_0=2.466\times10^{15}\,{\rm Hz}$)  in Figure~\ref{fig:mfactor},
where $\sigma_0 = 5.88\times10^{-14}\,{\rm cm^2}\, T_4^{-1/2}$  is the absorption 
cross-section for \Lya. The dust-free $M_{\rm F}$ may be fit  
from the Monte Carlo results (starred points in Figure~\ref{fig:mfactor}) as 
\begin{align}
M_{\rm F}^{\rm nodust} & \approx 29 \left(\frac{\tau_0}{10^{6}}\right)^{0.29} ~~~~ \left(\tau_0 \ge 10^{6}\right),
\label{eq:mf1}
\end{align}
\begin{align}
\log M_{\rm F}^{\rm nodust} \approx & -0.433 + 0.874 \, \log \tau_0 - 0.173\, \left(\log \tau_0\right)^2  \nonumber \\
                       & + 0.0133 \left(\log \tau_0\right)^3 ~~~~~~~~~~~~~~ \left(\tau_0 < 10^{6}\right)
\label{eq:mf2}
\end{align}
It can be seen that our measurements are largely consistent with the low temperature results from \citet{smith17}.
$M_F$ tends to be higher than those from the experiments with a higher temperature
\citep [$T=10^4\,{\rm K}$, ][]{bonilha79}, but this is expected, as  \Lya\ photons are trapped 
by the wing opacity, which is proportional to the Voigt parameter $a_V = 4.7\times10^{-4}\, T_4^{-1/2}$ \citep{adams72,adams75},
in the optically thick regime.

If dust is present and destroys \Lya\ photons, $M_{\rm F}$ cannot simply increase 
as $\propto \tau_{0}^{0.29}$ in the optically thick regime. Instead, one can expect that 
$M_{\rm F}$ saturates to a maximum value for a given optical depth, 
as the majority of \Lya\ photons are absorbed by dust. 
The right panel of Figure~\ref{fig:mfactor} demonstrates that in metal-rich, dusty 
environments, $M_{\rm F}$ cannot be more than $\approx 50$, whereas it can 
easily be as high as $\approx 300$ in metal-poor ($Z=0.01\,\Zsun$), 
less dusty ($D/M=0.04$) regions.

In order to properly model the momentum transfer from the scattering of \Lya\ photons, 
one should of course couple the \Lya\ MCRT to the RHD equations \citep[see][ for a one dimensional example]{smith17}.
However, this is not computationally feasible for three dimensional problems 
unless a specially designed algorithm, such as the discrete diffusion MC methodology \citep{smith17b}, is used.
 Instead, we introduce a sub-grid model that
can be used in our RHD simulations as follows. Since lower escape fractions of 
\Lya\ photons indicate that dust destroys \Lya\ photons more efficiently and 
limits the impact from \Lya\ pressure, we use the functional form of the escape 
fractions to find a fit that reasonably matches $M_{\rm F}$ in Fig.~\ref{fig:mfactor} 
(right panel).   In an optically thick slab, the escape fractions may be approximated 
as \citep{neufeld90,hansen06,verhamme06}, 
\begin{equation}
f_{\rm esc}^{\rm Ly\alpha} = 1 / \cosh \left\{ \frac{\sqrt{3}}{\pi^{5/12} \xi} [(a_V\tau_0)^{1/3} \tau_{\rm da}]^{1/2}   \right\},
\label{eq:fesc}
\end{equation}
where $\tau_{\rm da} = N_{\rm HI} f_{\rm d/m} Z'  \, \sigma_d (1 - \mathcal{A}_b)$, $\xi=0.525$ is a fitting parameter,
$Z'\equiv Z/Z_\odot$, and $f_{\rm d/m}$ is the dust-to-metal mass ratio, normalised to the value at solar metallicity.
Note that Eq.~\ref{eq:fesc} describes the fraction of \Lya\ photons that is eventually destroyed, hence
a simple combination of $M_{\rm F}^{\rm nodust}\,f_{\rm esc}^{\rm Ly\alpha}$ cannot be 
used as a proxy for $M_{\rm F}$ in dusty environments \citep[c.f.,][]{bithell90}. Instead, 
we note that the local maximum of $\left(M_{\rm F}^{\rm nodust}\, \times f_{\rm esc}^{\rm Ly{\alpha}}\right)$ 
gives a {\it minimum} estimate of the asymptotic value of $M_{\rm F,dust}$. We find this numerically 
by defining 
\begin{equation}
y \equiv \frac{\sqrt{3} }{\pi^{5/12} \xi } \left[ \left( a_V \, \tau_{\rm 0}\right)^{1/3} \tau_{\rm da}\right]^{1/2} .
\end{equation}
For the optical depth with a functional form of $M_{\rm F} \propto \tau_0^{0.29}$, the local maximum 
occurs at $y_{\rm peak}=0.71134$. Note that $M_{\rm F}$ is likely to be higher than this, 
because destroyed photons can also contribute to the total momentum before they get absorbed. 
To account for this, we modify the escape fractions by adjusting $\xi\rightarrow\xi_{\rm fit}=1.78$ 
so that it can reproduce the momentum transfer calculations from MCRT reasonably well 
(Figure~\ref{fig:mfactor}, right). The resulting optical depth at which 
$M_{\rm F}^{\rm nodust} f_{\rm esc}(\xi\rightarrow\xi_{\rm fit})\, $ reaches a peak may be written as
\begin{align}
\tau_0^{\rm peak} & = \left( \frac{\pi^{5/12} \xi_{\rm new}  }{ \sqrt{3}}\, y_{\rm peak} \right)^{3/2} \left[  \frac{ 1}{a_V^{1/3} (1-\mathcal{A}_b) f_{\rm d/m} Z'} \left( \frac{\sigma_0}{\sigma_d}\right) \right]^{3/4}  \nonumber \\
& = 4.06\times10^6 \, T_4^{-1/4}  \, \left( f_{\rm d/m}\, Z'\right)^{-3/4} \left( \frac{\sigma_{\rm d,-21}}{3}\right)^{-3/4}
\label{eq:tau0_peak}
\end{align}
Here $\sigma_{\rm d,-21}\equiv \sigma_{\rm d}/10^{-21}\,{\rm cm^2/H}$.

It is worth mentioning that the multiplication factor relies on the temperature of the 
medium \citep{adams75,smith17}. We perform the MCRT with a fixed temperature 
$T=100\,{\rm K}$, because this is the typical temperature of star-forming clouds 
under the influence of photoelectric heating on dust. If \Lya\ photons mostly reside
in the warm ISM ($T\sim10^4\,{\rm K}$), our model is likely to over-estimate 
the non-thermal pressure, although the dependence on temperature is not very strong ($\approxprop T^{-1/6}$) \citep{adams75}. 
However, it should be noted that warm gas exists mostly when ionizing radiation is 
 locally very intense or SN explosions are energetic enough to drive winds 
 at which point the \Lya\ force is no longer the only dominant mechanism. 
Therefore, we believe it is more appropriate to use the multiplication factor at low temperatures 
to investigate the impact of \Lya\ feedback.

\subsubsection{Subgrid implementation}
In practice, we include the momentum transfer from multi-scattered $\Lya$ photons as follows.
\begin{itemize}[leftmargin=0.3cm]
\item[1.] We first estimate the local $\Lya$ emissivity in every cells by computing the radiative recombination rates
\begin{equation}
\epsilon_{\rm rec,Ly\alpha} =  P_B (T)\, \alpha_B(T)\, n_e \, n_{\rm HII}\, e_{\rm Ly\alpha} ,
\label{eq:erec}
\end{equation}
where $e_{\rm Ly\alpha}$ is the energy of individual \Lya\ photon (10.16 eV), $n_e$ and $n_{\rm HII}$ are
the number density of electron and ionized hydrogen,  
$P_B(T) = 0.686 -0.106 \log T_4 - 0.009 \,T_4^{-0.44}$ is the probability for absorbed 
Lyman continuum photons to emit a \Lya\ photon \citep{cantalupo08}, and 
$\alpha_B$ is the case-B recombination coefficient \citep{hui97}, 
\begin{equation}
\alpha_B = 2.753\times10^{-14}~{\rm cm^3\,s^{-1}} \frac{\lambda^{1.5}}{\left[ 1+\left(\lambda/2.74 \right)^{0.407} \right]^{2.242}},
\end{equation}
where $\lambda = 315 614\,{\rm K} / T$. \Lya\ photons are also produced through
collisional excitation process \citep[e.g.,][]{callaway87}, but we find 
that this is a very minor contribution \citep[see also][]{rosdahl12,dijkstra14}. 
From Equation~\ref{eq:erec}, we compute the total number of \Lya\ photons produced 
in each cell per simulation time step, and take the maximum between this 
and the actual number of ionizing photons absorbed times $P_B(T)$, as
\begin{gather}
N_{\rm Ly\alpha} = \max\left[  \epsilon_{\rm rec,Ly\alpha} \Delta V \Delta t / e_{\rm Ly\alpha}, N_{\rm LyC}^{\rm abs} P_B(T)  \right],
\end{gather}
where $\Delta V$ is the volume of the cell, $\Delta t$ is the fine time step, and $N_{\rm LyC}^{\rm abs}$ is the 
total number of LyC photons absorbed during $\Delta t$.
The latter term is necessary, 
because the ionized fraction of hydrogen is often under-estimated in dense cells
where the Stromgren sphere is not fully resolved. 
\item[2.] Then, we estimate the multiplication factor ($M_F$). To do so, we compute the local 
column density using the Sobolev-type approximation \citep{gnedin09}, 
as $N_{\rm HII} \approx n_{\rm HI} l_{\rm Sob}$,
where $l_{\rm Sob} = \max(\rho /  | \nabla \rho|, \Delta x)$. 
Note that this should give a reasonable approximation as long as the optical 
depth ($\tau_0$) is large enough to give a value close to the maximum 
multiplication factor. 
Once we compute $\tau_0^{\rm peak}$ from Equation~\ref{eq:tau0_peak} and the corresponding 
 column density ($N_{\rm HI}^{\rm peak} = \tau_0^{\rm peak} / \sigma_0$), the multiplication factor can 
 be obtained from Equations~\ref{eq:mf1}--\ref{eq:fesc}, as
\begin{gather}
M_F=M_{F}^{\rm nodust}(\tau_{\rm 0}') \times f_{\rm esc}^{\rm Ly\alpha} (N_{\rm HI}', \xi_{\rm fit}),
\label{eq:mf_dust}
\end{gather}
where
\begin{gather}
\tau_0' = \min (\tau_0, \tau_0^{\rm peak})~~~~;~~~~N_{\rm HI'} = \min (N_{\rm HI}, N_{\rm HI}^{\rm peak}).
\end{gather}
\item[3.] We then impart a momentum of 
\begin{equation}
\Delta \vec{p} = M_{\rm F} \frac{N_{\rm Lya}\, e_{\rm Lya} }{c} \hat{n}
\end{equation}
using the flux-weighted average directions ($\hat{n}$) of ionizing photon groups\footnote{
Since the momentum transfer from \Lya\ photons in a homogeneous medium is expected to be 
spherically symmetric with respect to the stellar source, we adopt the directions of LyC photons,
which carry the information about the location of dominant stellar sources around the cell of interest,
as a simple approximation. }.
This method, however, cannot be used for the cell in which young star 
particles reside, because photons are emitted isotropically. Therefore, we take care 
of these cells with young stars ($t \le 50 \, {\rm Myr}$) separately, and inject the radial
momentum isotropically to 18 neighbouring cells, as is done for the 
transfer of momentum from SN explosions. Note that we limit the change in velocity 
to $|\Delta v| \le 100\,\kms$ per fine time step, because occasionally around low-density
ionization fronts, the velocity change is over-estimated due to our simple assumption 
of a static medium during acceleration\footnote{
In realistic situations, the multiplication factor would drop by an order of magnitude 
as the velocity of expanding shells increases to $100\,\kms$ \citep{dijkstra08}.
On the other hand, \Lya\ feedback is significant inside star-forming clouds 
where the escape velocity is on the order of $10\,\kms$. Indeed, we find that
the inclusion of \Lya\ feedback alone does not generate hot winds 
($T\gtrsim10^5\,{\rm K}$). We also confirm that limiting the velocity change to 
$|\Delta v| \le 30\, \kms$ does not make a significant difference.
}.
When momentum is cancelled out, we 
transfer kinetic energy to thermal energy.
We also note that this approach will neglect any \Lya\ photons 
escaping from their birth place (i.e. birth cell), which can potentially enhance the non-thermal pressure. 
We discuss this issue in Section 4.
\end{itemize}

\begin{figure}
   \centering
   \includegraphics[width=8.3cm]{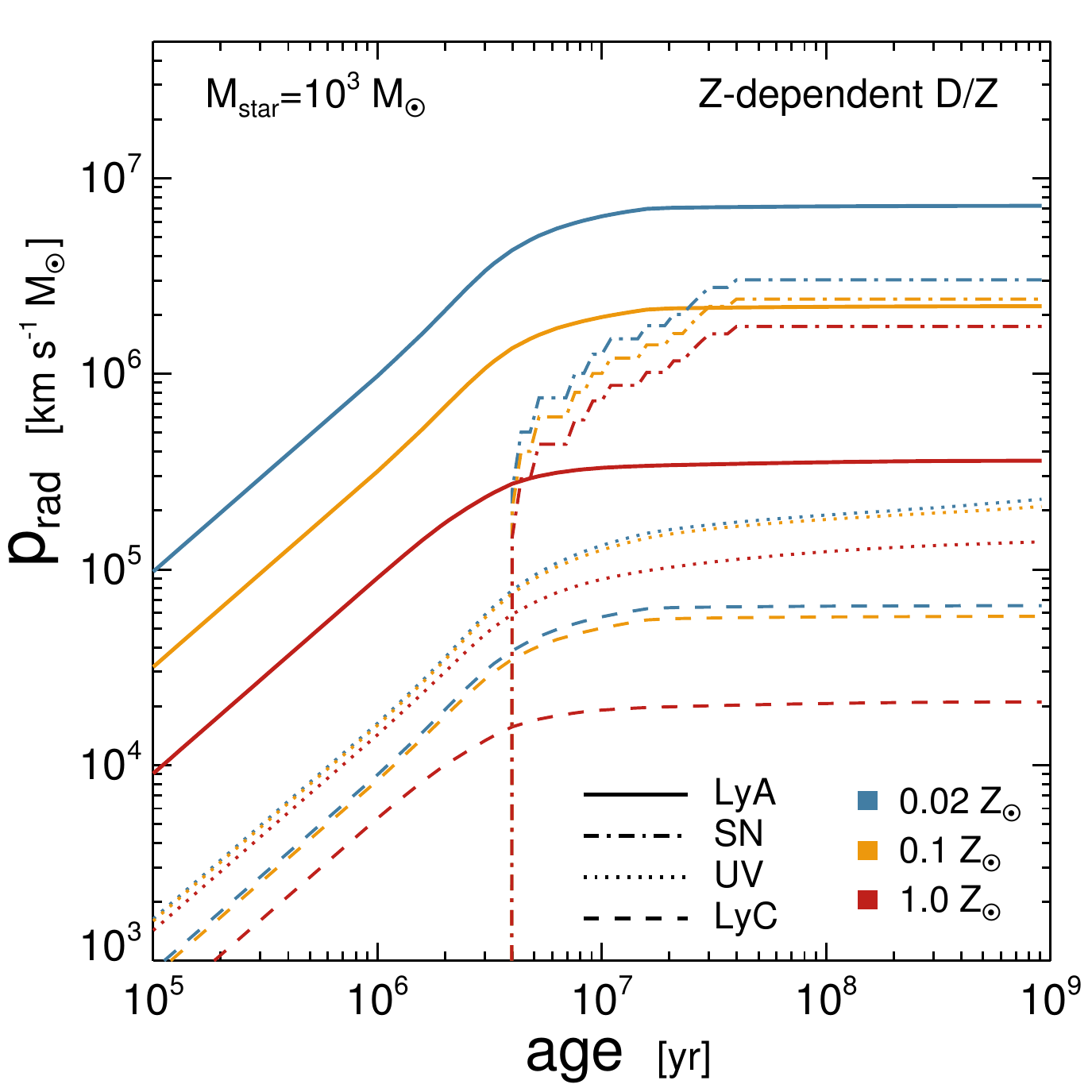} 
   \caption{ Integrated radial momentum from an SSP of $\mstar=10^3\,\msun$ 
   based on the BPASS spectra. The momentum budget from LyC ($\lambda<912\AA$, 
   dashed), UV ($\lambda <3000\AA$, dotted),  SN (dot-dashed), and \Lya\ (solid) is 
   shown with different line styles, as indicated in the legend. Different colour-codings 
   denote different metallicities.  To estimate the momentum from SNe, we randomly 
   sample the time delay for 11 SNe, appropriate for the Kroupa IMF, assuming 
   that they all explode in dense environments ($\nH=100\,\cmq$). Note that the 
   momentum would be augmented by a factor of $\sim 3$  if SN explodes at much 
   lower densities ($\nH=0.01\,\cmq$). The momentum from \Lya\ is estimated 
   assuming the metallicity-dependent dust-to-gas ratios (see text). One can see 
   that momentum transfer from \Lya\ pressure is significant particularly at low 
   metallicities.}
   \label{fig:mom}
\end{figure}

\begin{table*}
   \caption{Summary of idealised disk simulations embedded in a $10^{10}\,\msun$ 
   dark matter halo. From left to right, each column indicates 
   the name of the model, whether or not the simulations are performed with 
   on-the-fly radiative transfer, the minimum size of the computational cells, 
   the inclusion of mechanical SN feedback, \Lya\ pressure, photoionization 
   heating (PH), direct radiation pressure by UV (DP) and radiation pressure 
   by multi-scattered IR photons (IR), photoelectric heating on dust (PEH), 
   the metallicity of gas, and some remarks.}
   \centering
   \begin{tabular}{llllllllllll}
   \hline
  Model & RHD & $\Delta x_{\rm min}$ & $m_{\rm star}^{\rm min}$ & SN II & Lya & PH & DP & IR & PEH & Metallicity & Remarks\\
     \hline
     G8CO & -- & 4.6 pc & 910\,\msun & -- & -- & -- & -- & -- & -- &0.02 \Zsun &  \\
    G8SN & -- & 4.6 pc & 910\,\msun & \checkmark & -- & -- &--  & -- & -- & 0.02 \Zsun & \\
    G8R &\checkmark & 4.6 pc & 910\,\msun & -- & -- & \checkmark &\checkmark & \checkmark  & \checkmark & 0.02 \Zsun &  \\
    G8R-SN &\checkmark & 4.6 pc & 910\,\msun & \checkmark & -- & \checkmark &\checkmark  & \checkmark & \checkmark & 0.02 \Zsun & \\
    G8R-Lya &\checkmark & 4.6 pc &910\,\msun & -- & \checkmark &\checkmark &\checkmark  & \checkmark &  \checkmark & 0.02 \Zsun &  \\
    G8R-SN-Lya & \checkmark & 4.6 pc &910\,\msun & \checkmark &\checkmark & \checkmark   & \checkmark &\checkmark & \checkmark & 0.02 \Zsun  & Fiducial \\
        G8R-SN-Lya-f3 &\checkmark & 4.6 pc & 910\,\msun &\checkmark & \checkmark &\checkmark &\checkmark  & \checkmark &  \checkmark & 0.02 \Zsun & $L_{\rm star}\times3$  \\
    G8R-SN-Lya-f10 &\checkmark & 4.6 pc & 910\,\msun &\checkmark & \checkmark &\checkmark &\checkmark  & \checkmark &  \checkmark & 0.02 \Zsun & $L_{\rm star}\times10$  \\
    G8R-SN-Lya-s10 &\checkmark & 4.6 pc & 910\,\msun &\checkmark & \checkmark &\checkmark &\checkmark  & \checkmark &  \checkmark & 0.02 \Zsun & $\epsilon_{\rm ff}\times10$  \\
       \hline
   \end{tabular}
   \label{table:sim}
\end{table*}

\begin{figure*}
   \centering
   \includegraphics[width=17cm]{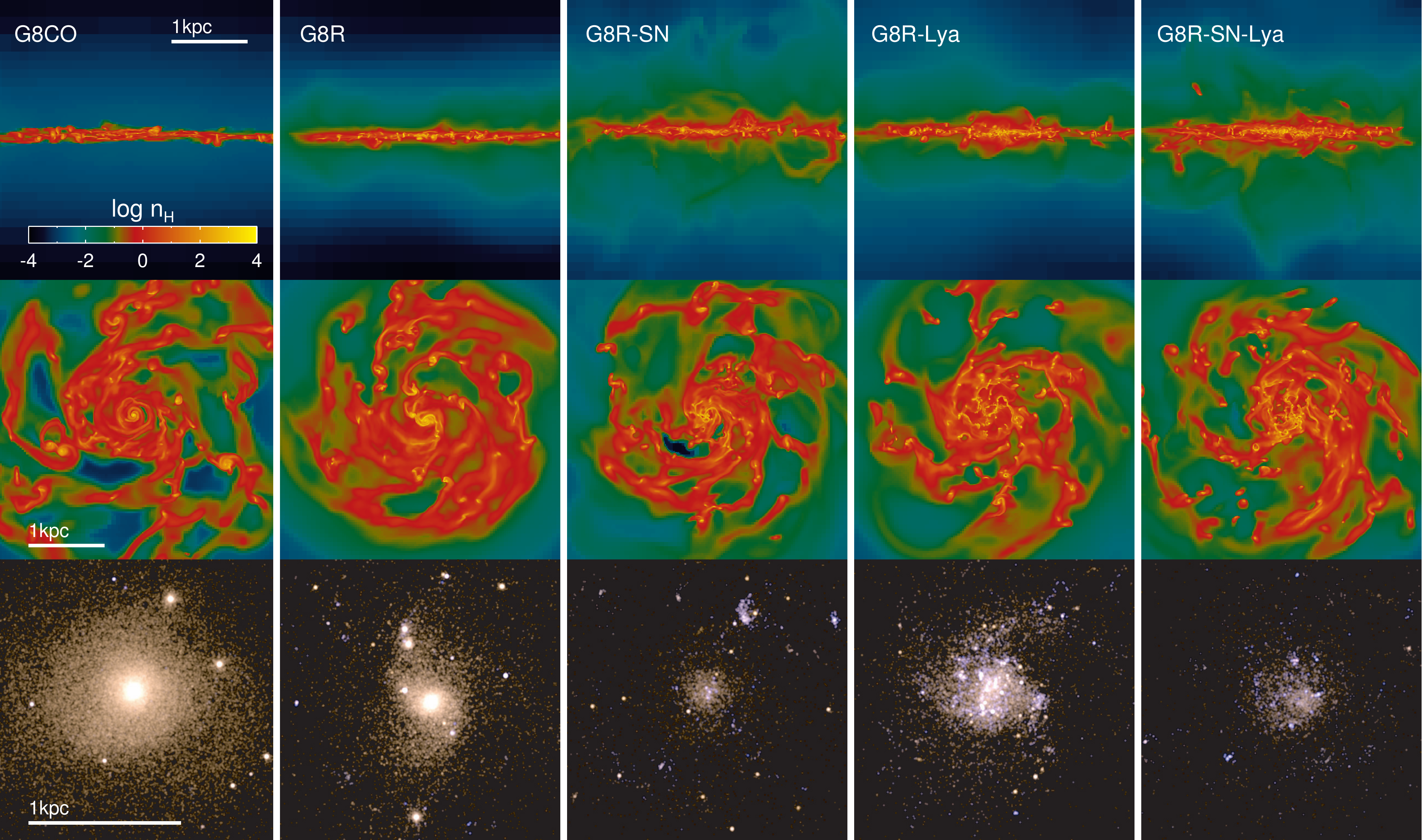}
   \caption{  
   Projected gas distributions of isolated disk simulations with different input physics 
   at $t=500\,{\rm Myr}$. The top and middle panels show the edge-on and face-on 
   view of the simulated galaxy, respectively. The bottom panels show the composite 
   stellar image using GALEX $NUV$, SDSS $g$,and $i$ bands. Note that a smaller 
   area is displayed for the stellar maps. Bluer colours correspond to younger stars. 
   The white scale bar denotes $1\,{\rm kpc}$.
   }
   \label{fig:pic}
\end{figure*}

In Figure~\ref{fig:mom}, we compare the relative importance of the momentum transfer 
from \Lya\ pressure with those from other processes in a metal-poor environment 
($Z_{\rm gas}=0.02 \,\Zsun$). We adopt the spectral energy distributions from the 
BPASS v2 model with the maximum stellar mass cut-off of $100\,\msun$ \citep{stanway16},
and assume that $\approx 68\%$ of LyC photons give rise to \Lya\ photons. 
Note that we use a higher metallicity grid ($Z=0.001$) for stars, as this is the 
lowest metallicity grid available in the BPASS v2 model.
We also consider the metallicity-dependent dust-to-metal mass ratio ($f_{\rm d/m}$),
normalised to the value at solar metallicity, in our simulations, 
based on the $X_{\rm CO,Z}$ case fitted with the broken power law from \citet{remy-ruyer14}, as
\begin{align}
\log f_{\rm d/m} & = 0 ~~ &  (x > 8.10), \nonumber \\
                         & = 1.25 -  2.10\times(x_\odot - x ) ~~ &  (x \le 8.10),
\label{eq:RR14}
\end{align}
where $x\equiv12 + \log \left(O/H\right)$ and $x_\odot=8.69$.
This means that the dust-to-metal ratio we adopt at $Z=0.02\,\Zsun$ is $\approx 200$ 
times smaller than the values usually found in the Milky Way \citep{draine07b,galametz11}.
For comparison, we also include the momentum input from LyC ($\lambda<912\AA$) 
and UV ($\lambda < 3000\AA$) photons. For the contribution from SNe, we randomly 
sample the delay time distributions using the method described in \citet{kimm15}, which is based on the SN II rates from {\sc starburst99} \citep{leitherer99}. 

Figure~\ref{fig:mom} illustrates that resonantly scattered \Lya\ in 
metal-poor ($Z \la 0.1\,\Zsun$), dense environments ($\nH\ge100\,\cmq$) 
can transfer as much radial momentum ($p_{\rm rad}$) as SNe even before 
massive stars ($>8\, \msun$)  evolve off the main sequence. 
 On the other hand, because \Lya\ photons are 
efficiently destroyed by dust, their pressure is likely to be sub-dominant compared to SNe 
in Milky Way-like galaxies or metal-rich, massive ($M_\star \ga 10^{10}\,\msun$) systems at high redshift \citep[e.g.][]{maiolino08}.
Nevertheless, it is worth noting that \Lya\ pressure can 
still drive substantially faster winds (by an order of magnitude), compared to 
direct radiation pressure from LyC or UV photons. One can also see that 
$p_{\rm rad}$ from \Lya\ has a stronger dependency on metallicity compared to that 
from UV photons, which is mainly due to the $Z$-dependent $f_{\rm d/m}$ we assume.

It is also useful to compare the effect of \Lya\ pressure with that of photo-ionization.
The maximum extent to which \Lya\ pressure can overcome 
the external pressure set by the ISM can be calculated as
\begin{align}
n_{\rm H} k_B T & = \frac{M_F L_\alpha}{4\pi r_\alpha^2 c} 
\end{align}
and thus
\begin{align}
r_{\alpha} & = \sqrt{\frac{M_{\rm F} L_{\alpha} }{4\,\pi c\, n_{\rm H} \,k_{\rm B}\, T }}
\label{eq:ralpha}
\\
                & \approx 37 \,{\rm pc} \,  \left(\frac{M_{\rm F}}{100}\right)^{1/2} \left(\frac{m_{\rm star}}{10^3 \,\msun}\right)^{1/2}    \left(\frac{P/k_{\rm B}} {10^5\,{\rm cm^{-3}\, K} } \right)^{-1/2} .\nonumber
\end{align}
This turns out to be larger than the maximum extent to which photo-ionization heating can
counter-balance the external pressure in most environments ($10^4 \le P/k_B \le 10^6\,{\rm cm^{-3}\, K}$) \citep{rosdahl15a},
\begin{align}
r_{\rm PH} & \approx 26 \,{\rm pc} \,  \left(\frac{m_{\rm star}}{10^3 \,\msun}\right)^{1/3}    \left(\frac{P/k_{\rm B}} {10^5\,{\rm cm^{-3}\, K} } \right)^{-3/2} \left(\frac{T_{\rm ion}}{10^4\,{\rm K}}\right)^{2/3}.
\label{eq:rph}
\end{align}
In particular, while $r_{\rm PH}$ decreases to sub parsec values if a star particle is deeply
embedded in a star formation cloud with $P/K\sim10^6\,{\rm cm^{-3}\,K}$, 
$r_{\alpha}\sim 20\,{\rm pc}$ still remains above our resolution. 
This means that including \Lya\ is also advantageous from a computational viewpoint,
as reasonably high resolution cells should be able to capture this process.
In addition, it would help to better resolve the Stromgren sphere near young stars 
by lowering the density into which LyC photons propagate.

\section{Results}

In this section, we examine the impact of radiation pressure from multi-scattered 
\Lya\ photons on galactic properties. For this purpose, we run seven RHD 
simulations of an isolated disk embedded in a $10^{10}\,\msun$ dark matter halo 
with different input physics, as outlined in Table~\ref{table:sim}. Also performed 
without the on-the-fly radiative transfer are \texttt{G8CO} and \texttt{G8SN} where 
we turn off local radiation and \Lya\ pressure. Thus, the uniform background UV 
radiation is the only feedback source in the former case, while SNe are the main 
energy source that governs the dynamics of the ISM in the latter. 
We use these models to isolate the effects of radiation feedback (photoionization 
heating, radiation pressure by UV and IR photons, and photoelectric heating on dust)
from other processes.

\subsection{Suppression of star formation}
We begin by investigating the suppression of star formation (SF) by different 
feedback processes. Figure~\ref{fig:pic} presents the projected gas distributions, 
and the composite images of SDSS u,g,i bands for stellar components. 
We generate these mock images by attenuating the stellar spectra using 
the method described in \citet{devriendt10}, which is based on the Milky way 
extinction curve \citep{cardelli89} and the empirical calibration by \citet{guiderdoni87}.

\begin{figure}
   \centering
   \includegraphics[width=8.5cm]{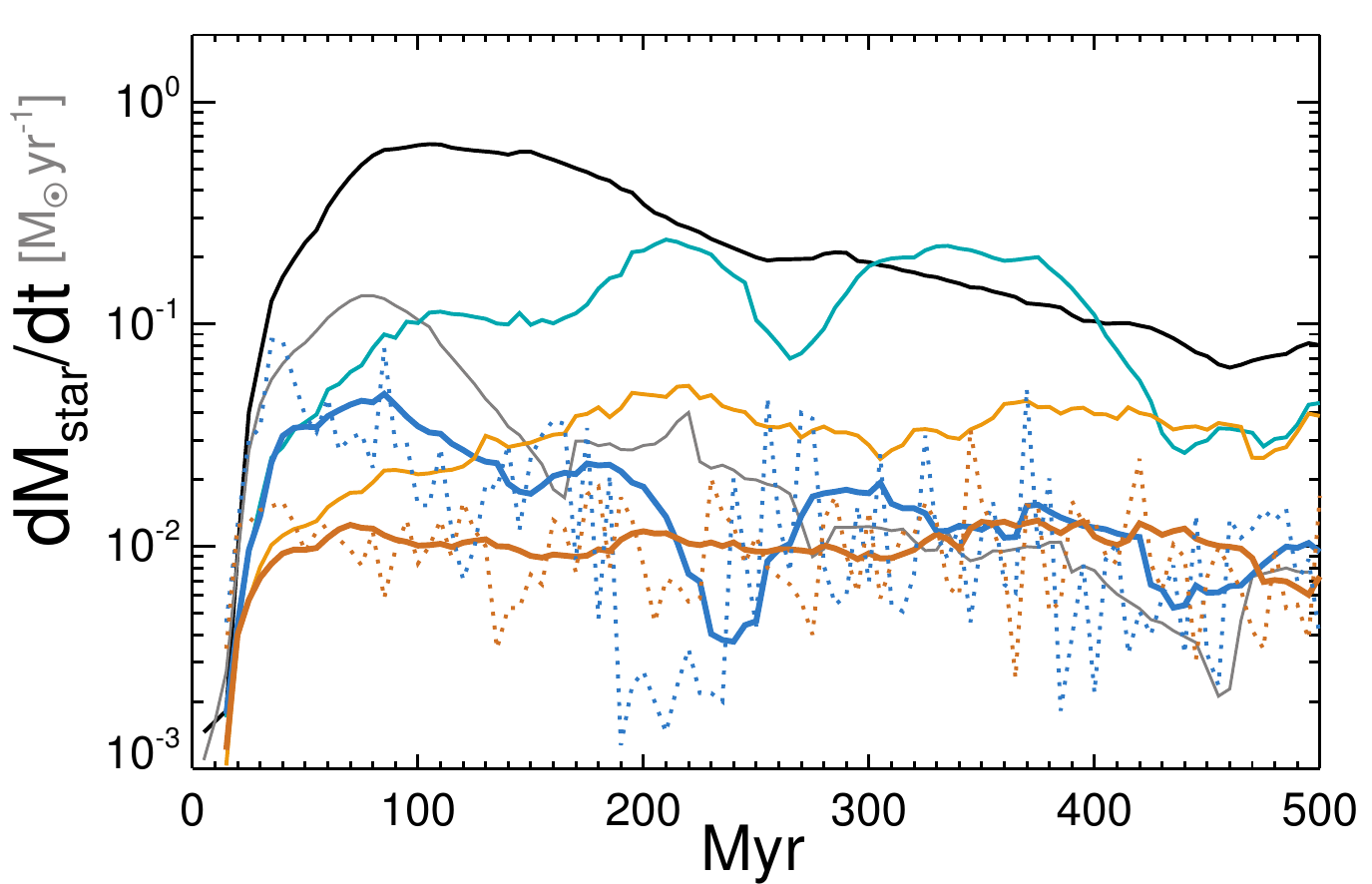}
    \includegraphics[width=8.5cm]{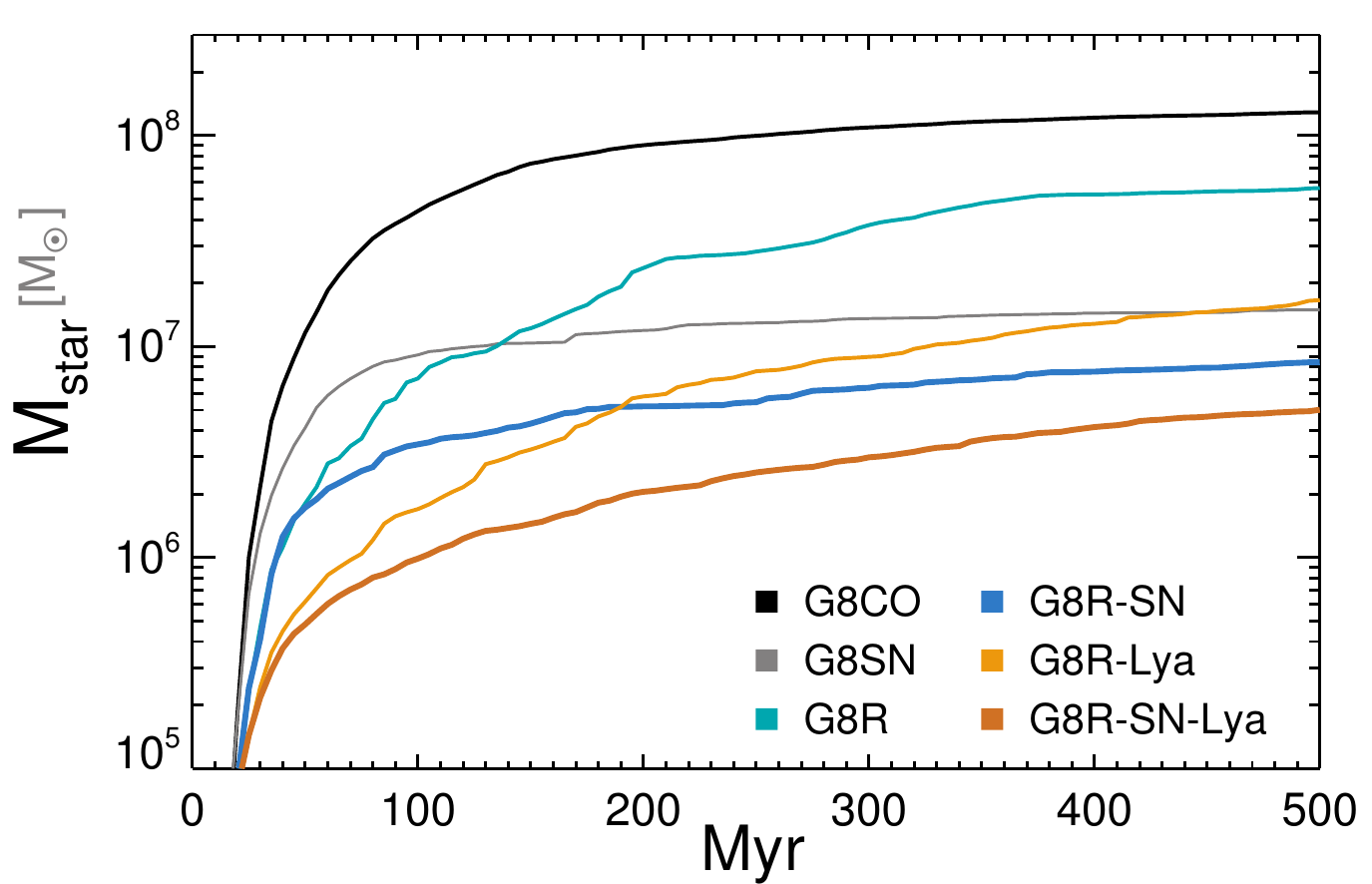}
   \caption{  
{\it Top panel}: Star formation histories of the simulated galaxy with 
different feedback processes, as indicated in the legend. We average the star
formation rates over 50 Myr to make the comparison easier (solid lines). 
Also included as dotted lines are the star formation rates averaged over shorter 
timescale (5 Myr) for \texttt{G8R-SN} and \texttt{G8R-SN-Lya} models. Note that 
star formation based on the thermo-turbulent model is generally bursty.  
{\it Bottom panel}: The integrated stellar mass formed as a function of time. 
The inclusion of \Lya\ pressure regulates the star formation in the early phase, 
and reduces the total stellar mass by a factor of two compared to the run without it. }
   \label{fig:sf}
\end{figure}

The run without any feedback source (\texttt{G8CO}) produces a massive stellar 
core at the galactic centre, which is usually found in simulated galaxies suffering 
from artificial radiative losses \citep[e.g.][]{katz92,agertz13,hopkins14,kimm15}.
Even the model with the three well-known radiation feedback processes 
(photo-ionization heating + radiation pressure by UV and IR photons, \texttt{G8R}) 
cannot prevent the formation of the massive stellar core. The insignificance of radiation 
pressure by UV photons is not surprising, given that the momentum budget is not 
substantial \citep[e.g.][]{leitherer99,kimm17}. IR pressure is not strong either, 
because the optical depth to IR photons is small in these types of metal-poor systems 
\citep{hopkins12a,rosdahl15b}. The impact of photoelectric heating on dust is also 
 expected to be minimal because the amount of dust is negligible, 
 although the star-forming regions are heated to $\sim 100\,{\rm K}$ from 
 $\sim 10\,{\rm K}$ \citep[see also][c.f. \citealt{forbes16}]{hu17}. 
In contrast, photoionization heating exerts extra pressure on the ISM 
\citep{krumholz07b,dale12,gavagnin17}, delaying runaway gas collapse \citep{rosdahl15b}. 
This is evident by comparing the two runs, \texttt{G8SN} and \texttt{G8R-SN}, 
where purely hydrodynamic simulations form a lot more stars in the early phase.
Figure~\ref{fig:sf} shows that the stellar mass is reduced by a factor 
of $\sim2$ at 500 Myr when radiation feedback is included.

\begin{figure}
   \centering
     \includegraphics[width=7.5cm]{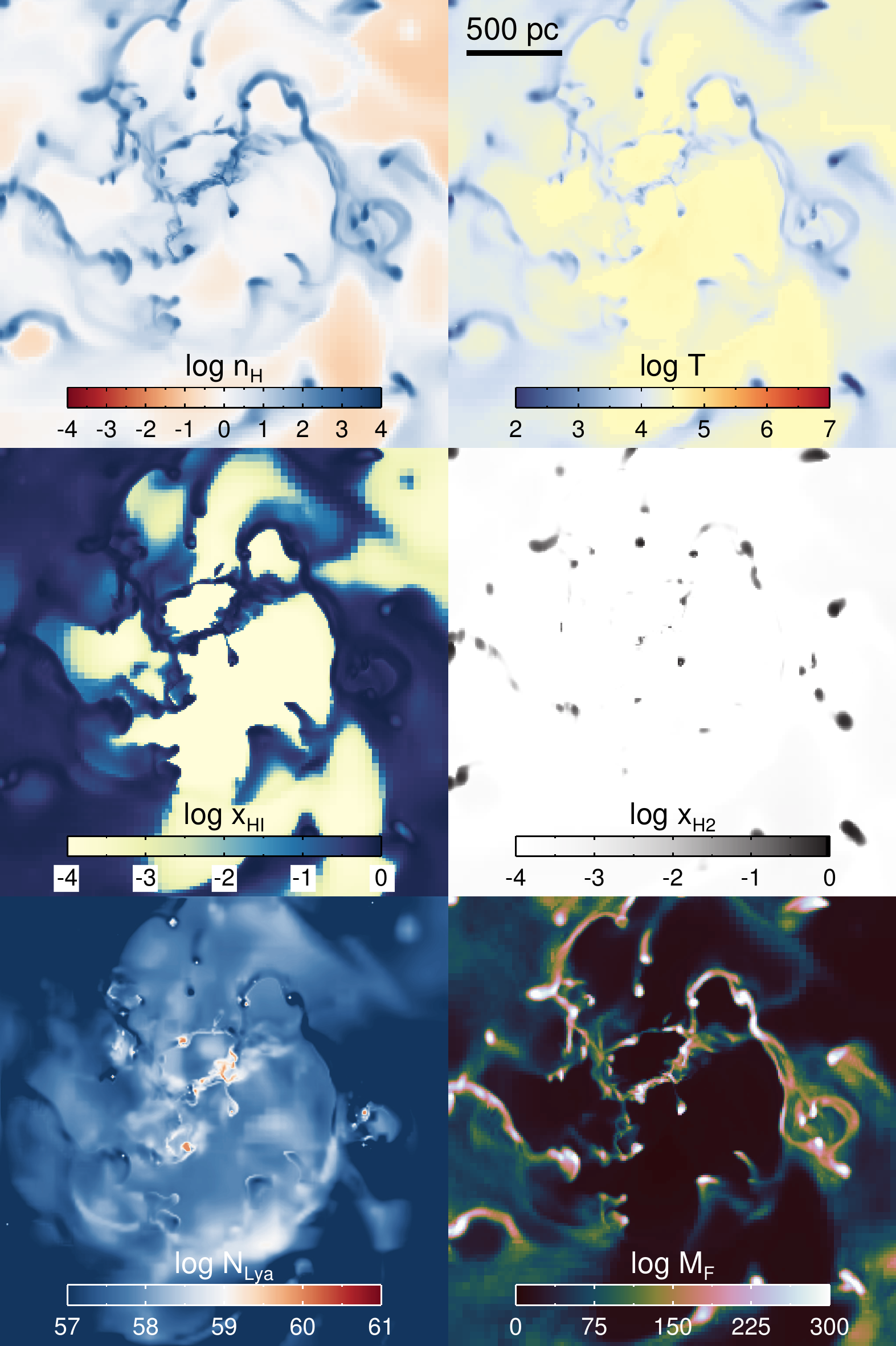}
   \caption{  
An example of the projected distribution of $\Lya$ multiplication factor 
in the central region of the \texttt{G8R-Lya} run at $t=500\,{\rm Myr}$.
The panels show the density, temperature, neutral hydrogen fraction ($x_{\rm HI}$), 
molecular hydrogen fraction ($x_{\rm H_2}$), 
number of \Lya\ photons produced in each cell ($N_{\rm Lya}$),
and the multiplication factor ($M_F$).
\label{fig:pic_ex}
   }
\end{figure}

More interestingly, we find that \Lya\ pressure can suppress SF further by a factor 
of $\sim 5$ compared to the run with three well-known radiation processes (\texttt{G8R} vs \texttt{G8R-Lya}, Figure~\ref{fig:sf}). The dense stellar core found in the 
\texttt{G8R} run no longer exists (Figure~\ref{fig:pic}), as the massive gas cloud 
at the galactic centre is efficiently dispersed. This is more evident in 
Figure~\ref{fig:pic_ex} where we plot the central region of the \texttt{G8R-Lya} run.
The size of molecular clouds are significantly smaller ($l \la 100 \,{\rm pc}$) than the 
massive cloud found in \texttt{G8CO} or \texttt{G8R} ($\sim200-500 \,{\rm pc}$) 
(Figure~\ref{fig:pic}). It is worth noting that \Lya\ pressure operates essentially 
on star-forming cloud scales, as the number of LyC photons that generate \Lya\ 
photons falls precipitously as $\propto d^{-2}$, where $d$ is the distance from the source
(Figure~\ref{fig:pic_ex}). 
Moreover, since scattering of \Lya\ requires neutral hydrogen, this mechanism is 
effective only in the cold ISM. The multiplication factor is close to zero in the 
ionized, warm ISM, whereas it is maximised ($\sim300$)
in the cold dense phase. Note that in this figure we compute $M_F$ by counting 
molecular hydrogen as part of the neutral hydrogen, because it is readily destroyed 
by Lyman-Werner radiation near hot stars.
 
As discussed in Figure~\ref{fig:mom}, \Lya\ pressure plays an important role even 
before the onset of SN explosions. The \texttt{G8R-Lya} run shows a significantly less SF than the \texttt{G8R-SN} run during the initial collapsing phase (Figure~\ref{fig:sf}). 
The overall SF histories also appear more smooth 
in the presence of \Lya\ pressure, as the feedback comes into play during the early
evolution of individual clouds and prevents too many stars from forming.
This process is quite different from the way SNe work in the sense that a big burst of SF 
leads to strong outflows, which in turn suppresses SF, leaving the histories bursty.

\begin{figure}
   \centering
     \includegraphics[width=4.1cm]{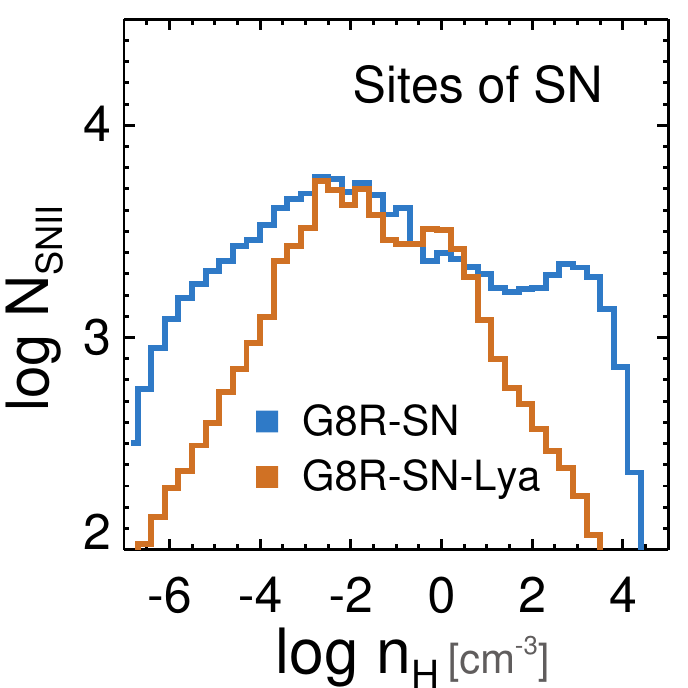}
     \includegraphics[width=4.1cm]{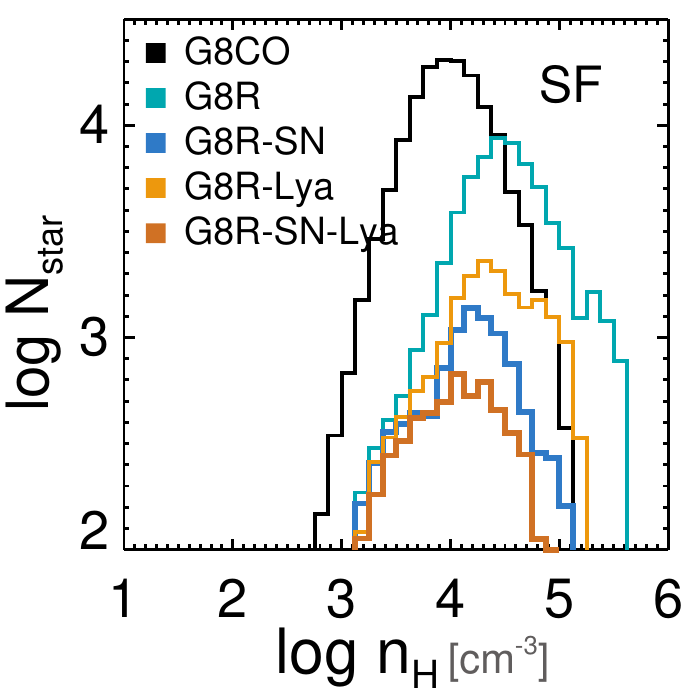}
     \includegraphics[width=4.1cm]{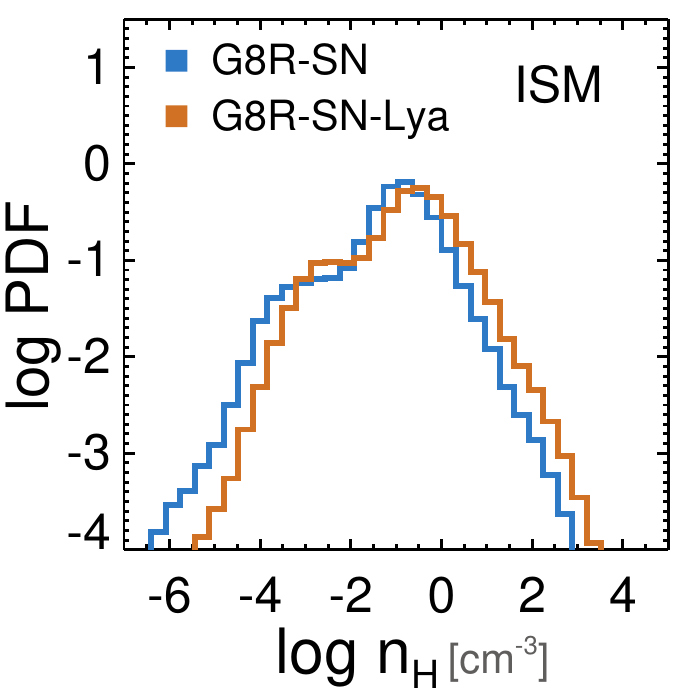}
      \includegraphics[width=4.1cm]{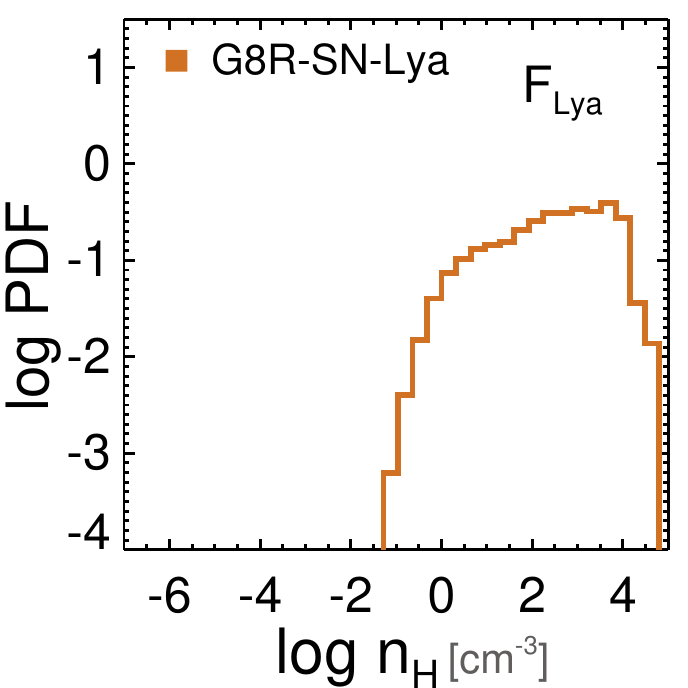}
   \caption{  
Distributions of densities at which SN explodes (top left) and star particles are 
formed (top right).  The dense gas ($\nH\ga100\,\cmq$) is efficiently dispersed 
before the onset of SNe in the presence of \Lya\ pressure in the metal-poor 
system. The fraction of star particles formed at very high densities ($\nH\ga10^4\,\cmq$) is also reduced.
The bottom left panel shows the volume-weighted density distributions of the ISM around the central mid-plane  
($\sqrt{x^2+y^2} < r_{\rm half,m}$ and $|z| < H$), where $r_{\rm half,m}$ is the half-mass radius and $H$ 
is the scaleheight (see text). The bottom right panel indicates the volume-weighted densities at which 
\Lya\ force ($\propto N_{\rm Lya} M_F$) is significant.
}
   \label{fig:snsites}
\end{figure}

Although the radial momentum imparted from $\Lya$ can be more significant 
than that from SNe in dense regions (Figure~\ref{fig:mom}),
we find that the regulation of SF is slightly more efficient in the run 
with SNe (\texttt{G8R-SN}) than in the run with \Lya\ (\texttt{G8R-Lya}). 
This is because a large fraction of SNe explode in low density environments 
($\nH\sim10^{-2}\,\cmq$) due to spatially correlated explosions from a realistic 
time delay in our dwarf-sized galaxy 
\citep[Figure~\ref{fig:snsites},][]{kimm15}. Since the momentum input from SNe 
is dependent on the ambient density ($p_{\rm SN}\propto \nH^{-2/17}$), 
it is enhanced by a factor of three, compared to the explosions occurring inside a GMC 
which has a mean density, $\nH\sim100\,\cmq$. Moreover, while SNe are an 
impulsive process, \Lya\ pressure imparts momentum continuously over several Myrs. 
Thus, the radial momentum from \Lya\ may be counter-balanced by ram pressure 
and self-gravity of clouds that act on a free-fall timescale of a few Myr. 

The stellar mass of the simulated galaxy is further reduced by 40\%, compared to 
the \texttt{G8R-SN} case, when both \Lya\ and SN feedback are included 
(\texttt{G8R-SN-Lya}). Most of the suppression occurs  during the early phase ($t\la 200\,{\rm Myr}$)
outside the half-mass radius  
($r_{\rm eff,m}=0.44\,{\rm kpc}$) where star clusters form sporadically.
The total mass formed outside $r=0.5\, {\rm kpc}$ over 500 Myr period is 
$M_{\rm form, out}=2.1\times10^6\,\msun$ (\texttt{G8R-SN-Lya}), 
whereas it is roughly twice greater ($M_{\rm form,out}=5.0\times10^6\,\msun$)
in \texttt{G8R-SN}. In contrast, a similar amount of stars are formed in the 
central region ($r\le0.5\, {\rm kpc}$) ($M_{\rm form,in}=3.4\times10^6\,\msun$ vs. 
$M_{\rm form, in}=2.9\times10^6\,\msun$). This happens because 
 there is a larger amount of gas in the inner region in \texttt{G8R-SN-Lya} 
($M_{\rm gas, in}^{\rm no Lya}\sim2\times10^7\,\msun$) than in 
\texttt{G8R-SN} ($M_{\rm gas, in}^{\rm Lya}\sim10^7\,\msun$).  The gas is less efficiently blown 
away from the galaxy due to a less bursty star formation history
and falls back to the central region (see the next section).
As a result, the difference in stellar mass becomes less prominent at late 
times.\footnote{ This implies that the suppression due to \Lya\ pressure might 
not be as significant  as found in this study if gas accretion onto metal-poor galaxies 
occurs smoothly at high redshifts \citep[c.f.][]{kimm14,hopkins14,wise14}.}

\citet{hennebelle14,walch15} demonstrate that the choice of SN driving, 
i.e. where the SN explosions should be placed, leads to vastly different star 
formation histories. This is because SN bubbles cannot propagate efficiently 
if they explode in dense pockets of gas \citep{iffrig15}. Thus, understanding the local 
conditions for SNe is an important step towards a complete picture of the evolution 
of the ISM. Figure~\ref{fig:snsites} (top left panel) shows that the number of SNe exploding 
in dense environments is very sensitive to the radiation feedback. 
More diffuse gas tend to surround young massive stars in the \texttt{G8R-SN-Lya} run, 
whereas still some number of SNe explode in dense regions in 
\texttt{G8R-SN} ($\nH\ge100\,\cmq$). The inclusion of \Lya\ photons also reduces 
the number of SNe exploding in low-density regions ($\nH\la10^{-3}\,\cmq$),
as SNe become less correlated. Given that the typical density at which SNe 
explode better matches that of volume-filling gas  (bottom left) than that of the star-forming 
regions (top right), our results suggest that the random SN 
driving model, which leads to the most significant outflows \citep{walch15}, may be 
the most relevant situation under strong radiation pressure. We also find that the number of stars born in 
a very dense medium ($\nH\ga10^4\,\cmq$) is decreased (top right) in the  \texttt{G8R-SN-Lya} run,
as momentum change due to \Lya\ pressure ($\propto N_{\rm Lya} M_F$) 
is strong at high densities (bottom right). We can confirm 
that the local virial parameter in the star-forming regions is increased as well, 
indicating that the star-forming gas becomes less gravitationally bound.

\begin{figure}
   \centering
     \includegraphics[width=8.5cm]{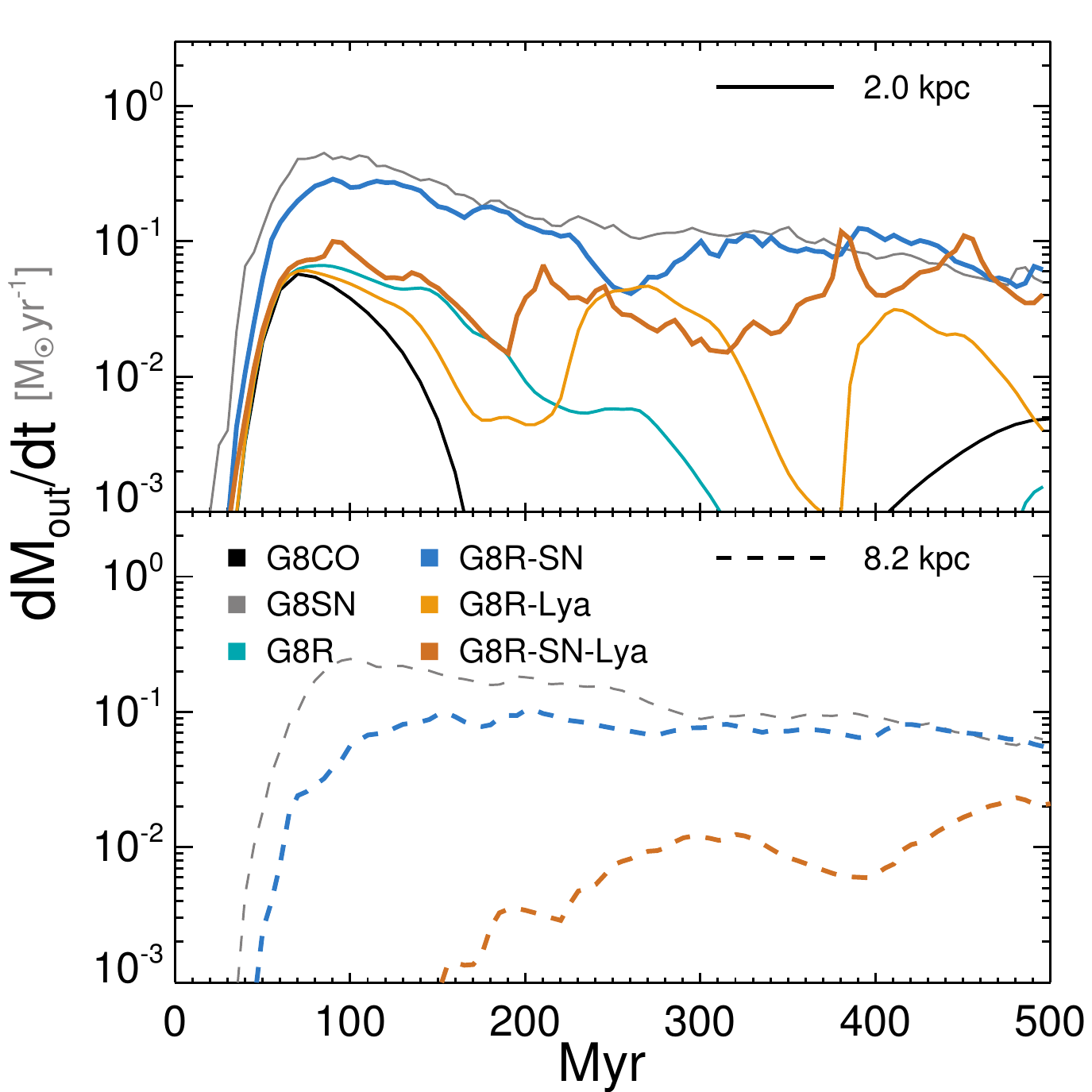}
   \caption{ Galactic outflow rates measured at two different heights 
   ($|z|=2\,{\rm kpc}$; top panel,  $|z|=8.2\, {\rm kpc}$: bottom panel). 
   Different models are shown as different colour-codings, as indicated 
   in the legend.  The models with weak feedback (NoFB, R, and R-Lya) 
   cannot launch strong outflows that reach $0.2\,R_{\rm vir}$ ($=8.2\,{\rm kpc}$), 
   and thus does not appear in the bottom panel. 
  }
   \label{fig:out}
\end{figure}

\begin{figure}
   \centering
     \includegraphics[width=8.5cm]{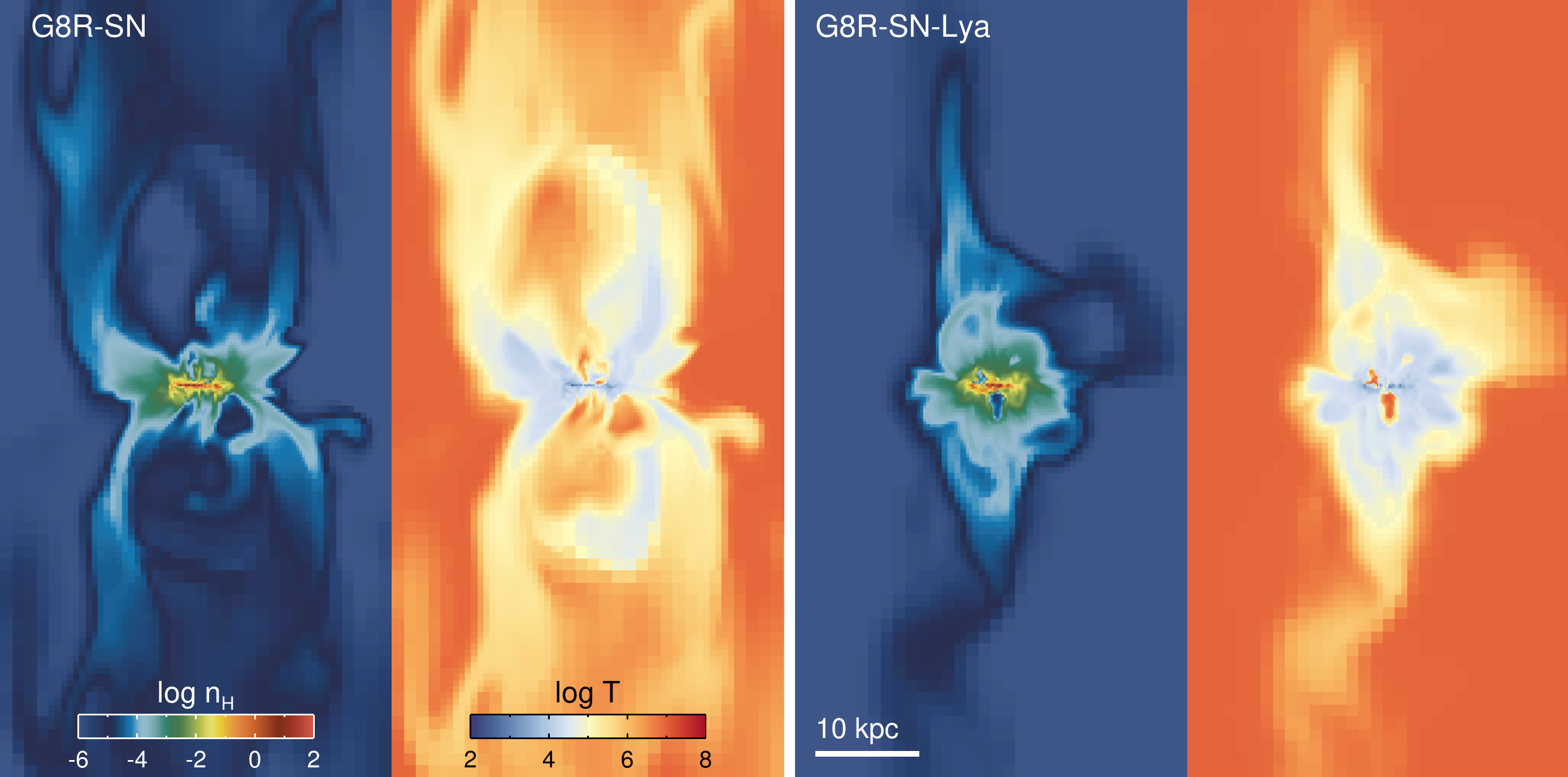}
   \caption{ Slices of density and temperature distributions from the two 
   runs (\texttt{G8R-SN}: left, \texttt{G8R-SN-Lya}: right) at $t=500\,{\rm Myr}$. 
   Each plot measures $37.5\,{\rm kpc}\times75\,{\rm kpc}$.   }
   \label{fig:out_ex}
\end{figure}

\begin{figure}
   \centering
     \includegraphics[width=8.5cm]{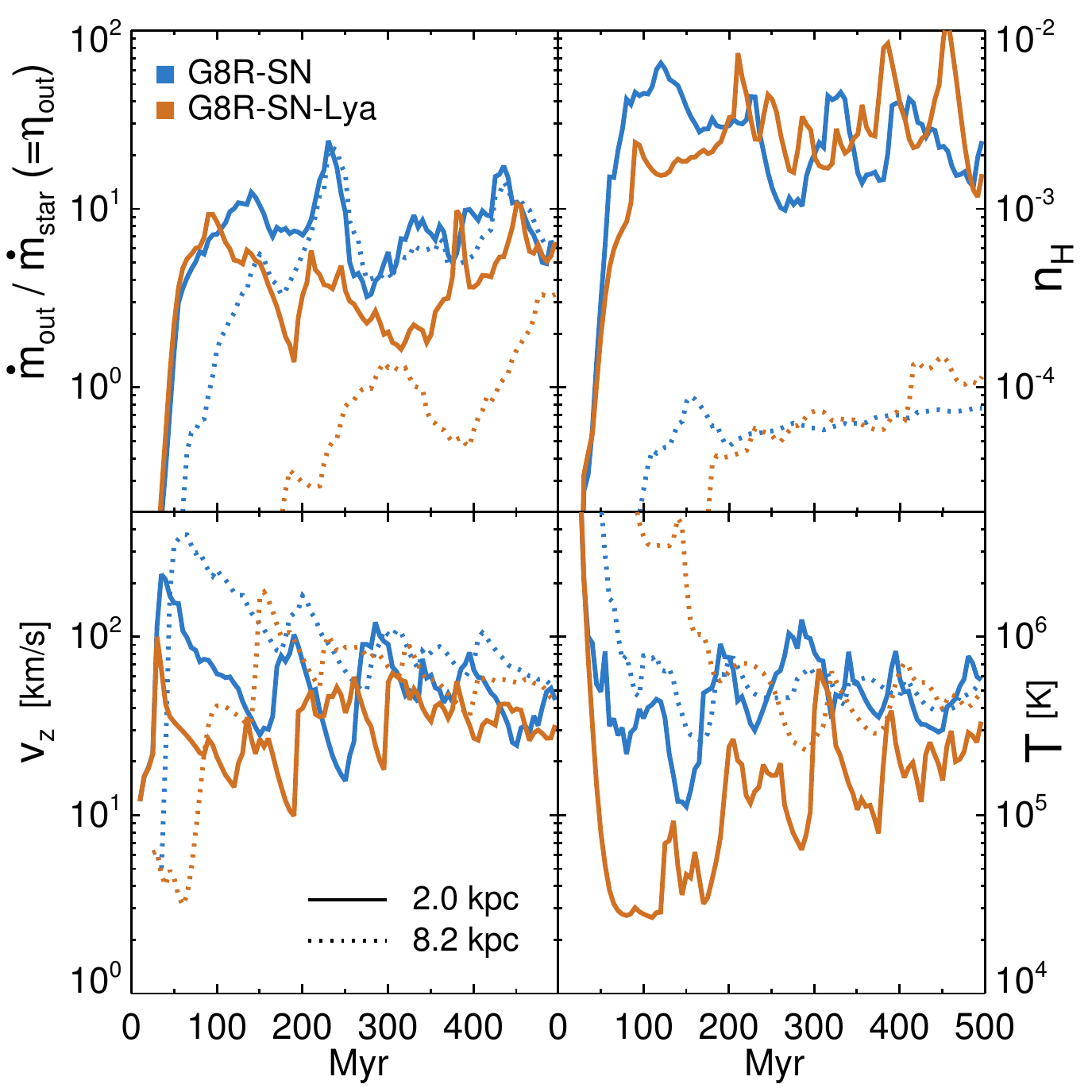}
   \caption{ Properties of the galactic outflows in the runs without (\texttt{G8R-SN})
  and with \Lya\ pressure (\texttt{G8R-SN-Lya}). From the top left, each panel shows 
  the outflow rates in $\msun\,{\rm yr}^{-1}$, flux-averaged density in $\cmq$, 
  outflow velocities in $\kms$, and temperature in Kelvin. The solid lines denote 
  the properties measured at $|z|=2\,{\rm kpc}$, whereas the properties measured
  at $|z|=8.2\,{\rm kpc}$ are shown as the dotted lines. Outflows tend to be slower 
  and cooler in the run with \Lya\ pressure.
   }
   \label{fig:outdetail}
\end{figure}

\subsection{Galactic outflows}
Galactic outflow rates are a useful measure of the strength of stellar 
feedback. The results from cosmological simulations suggest that high-$z$ 
dwarf galaxies should host strong outflows with an $\dot{m}_{\rm out}$ greater in magnitude than the
star formation rates in order to reproduce the observationally derived stellar 
mass, rotation curves, and gas metallicities in dwarf-sized galaxies 
\citep{finlator08,kimm15,muratov15}. Local observations of small starburst 
systems \citep{heckman15,chisholm17} also seem to suggest a mass-loading 
factor greater than unity ($\eta_{\rm out}\ga1$), which is defined as the ratio of 
mass outflow and star formation rates.

In Figure~\ref{fig:out}, we show the mass outflow rates of our simulated galaxies, which 
are measured by computing the mass flux at $|z|=2\,{\rm kpc}$ and 
$|z|=8.2\,{\rm kpc}$ $(=0.2\,R_{\rm vir})$. Since the initial gas distributions are 
not under hydrostatic equilibrium, some gas leaves the mid plane, leading to an 
initial outflow. This is the major cause of the outflows before $t\la150\,{\rm Myr}$ 
in the runs with very weak feedback (\texttt{G8CO} and \texttt{G8R}), although 
cloud-cloud interactions occasionally give rise to outflowing events 
(e.g., $t\approx$ 400-500 Myr).

We find that the outflows are strongest in the \texttt{G8R-SN} and \texttt{G8SN} 
runs. These models show a similar mass flux at the two different heights 
($|z|=2$ and $8.2\,{\rm kpc}$), which indicates that SNe drive fast, 
mass-conserving winds that propagate out to the CGM. At $t=500\,{\rm Myr}$, 
a bipolar outflow extends to the virial radius of the dark matter halo 
$\sim40~{\rm kpc}$ (Figure~\ref{fig:out_ex}). Surprisingly, we find that the 
addition of \Lya\ pressure leads to weaker outflows than \texttt{G8R-SN}. 
Even though the two runs sometimes display comparable outflow rates 
in the inner region ($|z|=2\,{\rm kpc}$), the large-scale bipolar outflows are 
not observed in the run with \Lya\ pressure. 

To better understand the properties of outflows from these simulations, 
we measure the mass-loading factors, the typical density, velocity, and temperature 
of the outflows responsible for most of outflowing mass in Figure~\ref{fig:outdetail}. 
To do this, we average the latter quantities by weighting the mass flux 
($\rho v_r \Delta x^2$). To estimate the mass-loading factor, we use the star 
formation rates averaged over 50 Myr, as the instantaneous rates vary widely 
in time. Several interesting features can be inferred from this plot. 
First, the mass-loading factor in the \texttt{G8R-SN} run is larger than \texttt{G8R-SN-Lya}. 
The time average of the mass-loading factor measured at $|z|=2\,{\rm kpc}$ 
during a settled period ($300 \le t \le 500\,{\rm Myr}$) is 
$\eta_{\rm out}=8.0^{+3.1}_{-2.2}$ and $4.0^{+2.9}_{-1.7}$ in the \texttt{G8R-SN} 
and \texttt{G8R-SN-Lya} run, respectively. Note that the mass loading without
 \Lya\ photons ( \texttt{G8R-SN}) is broadly consistent with the independent work by \citet{hu17}.
The difference in the mass-loading between the two models becomes more 
pronounced if we compare the loading factors at $8.2\,{\rm kpc}\, (=0.2\, R_{\rm vir}) $ 
($\eta_{\rm out}=6.8$ vs $1.0$). This is indicative that the powerful outflows seen 
in the SN run is not solely due to higher {\it average} star formation rates.

The velocity of the outflows in the inner region of the galaxy tends to be faster in the run 
without \Lya\footnote{ Note that the outflow 
velocity in the run with \Lya\ could increase further if the gas were irradiated by a non-thermal Compton-thick spectrum from massive 
black holes, as the harder spectrum allows for neutral hydrogen to survive longer than the normal Pop II spectrum  \citep{smith16}. } , compared to the fiducial run.
The inner galactic winds often have 
$v_z \ga 100\,\kms$, which is larger than the escape velocity of the halo 
($v_{\rm esc}\sim40\,\kms$). On the other hand, a more gentle velocity of 
$v_z \sim 20$--$60\,\kms$ is seen in our preferred model (\texttt{G8R-SN-Lya}). Similarly, the temperature 
of the inner outflows is also cooler in the run with \Lya\ 
($T\sim 0.5\times10^5$ -- $3\times10^5\,{\rm K}$), even though the typical density 
responsible for the outflows are comparable. 
As the winds propagate outwards, some of the component with low velocity falls back and
does not contribute to the outflows in the outer region of the halo, leading to the slight 
increase in velocity at $r=0.2\,\rvir$. This is notable in the \texttt{G8R-SN-Lya} run  
where the temperature of the outflows is increased to $\sim4\times10^5\,{\rm K}$ 
in the outer region of the halo.

We also find that the density of the winds varies substantially as a function of radius. 
The density of the outflows at $|z|=2\, {\rm kpc}$ is $\nH\sim0.003\,\cmq$, 
while it is reduced by more than an order of magnitude at $|z|=8.2\, {\rm kpc}$. 
We mainly attribute this to the fact that the internal pressure of the outflows 
($P/k_B\sim 10^3\, \cmq\,{\rm K}$) is larger than the pressure in the ambient medium 
($P/k_B=10\, \cmq\,{\rm K}$).  For comparison, the ram pressure exerted by the halo 
gas is negligible for the initial conditions chosen ($P_{\rm ram}/k_B\sim 1\,\cmq\,{\rm K}$). 
Thus, the reduction in density is likely caused by adiabatic expansion. However, 
we note that our refinement strategy may also be partially responsible for the lower density. 
Since refinement is triggered based on mass and jeans length (i.e. not on pressure), 
computational grids are split slower than the propagation of outflows, which may have 
resulted in more diffuse structures that it should. 

\begin{figure}
   \centering
  \includegraphics[width=8.5cm]{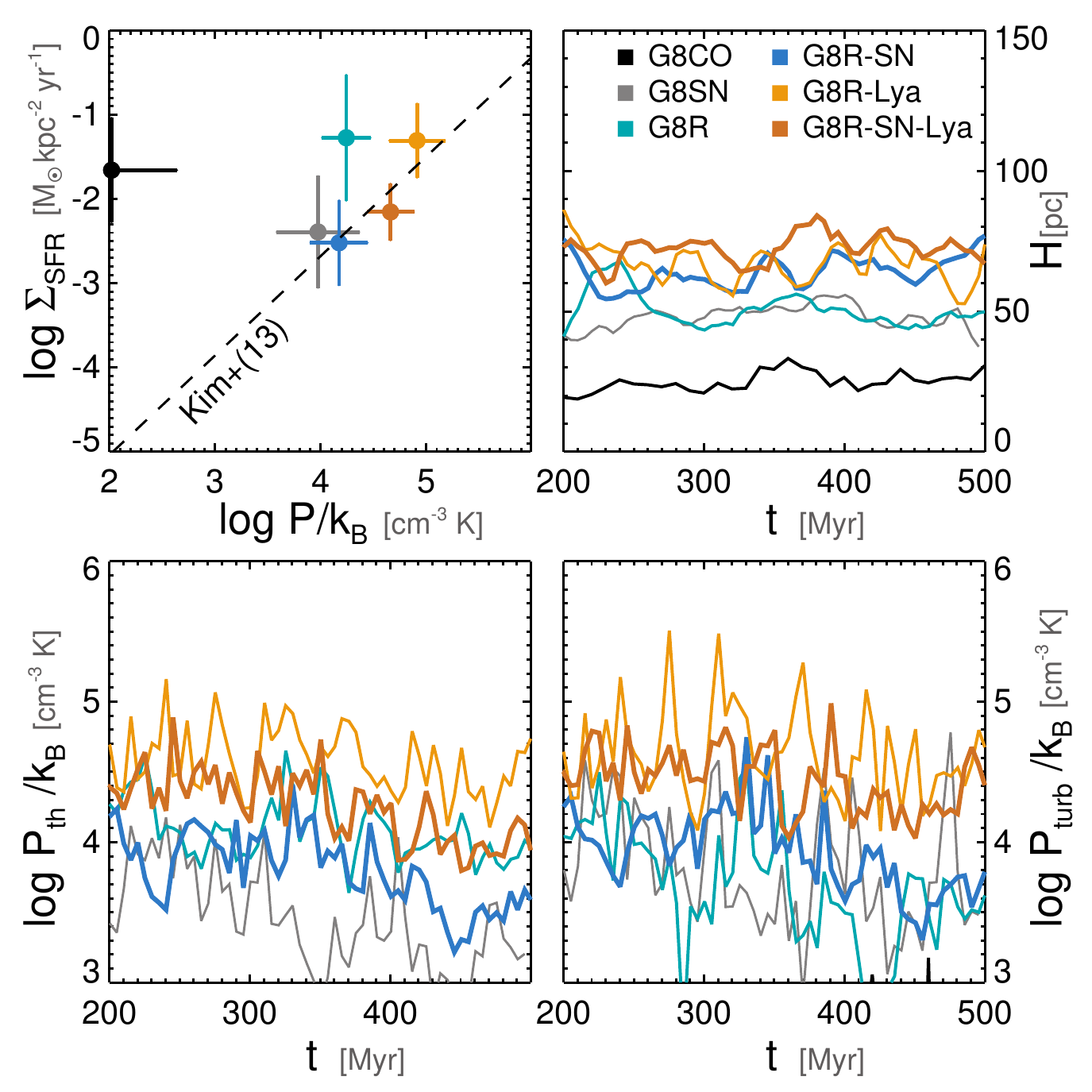}
   \caption{ Effects of different feedback processes on the vertical structure of the ISM. 
   The top left panel shows the relationship between the total pressure (turbulent + 
   thermal)  and the star formation rate density ($\Sigma_{\rm SFR}$). The dashed line 
   displays the fit to the results from local shearing box simulations by \citet{kim13}. 
   The top right, bottom left, and bottom right panel presents the time evolution of the 
   scale-height of the gaseous disk, thermal pressure, and turbulent pressure in the mid 
   plane ($|z|\le H$). It can be seen that the disk becomes thicker in the run with 
   \Lya\ feedback (\texttt{R-SN-Lya}) than the run without it (\texttt{R-SN}).
    }
   \label{fig:ISM}
\end{figure}

\subsection{Structure of the Interstellar medium}

Small-scale simulations demonstrate that feedback from SNe is essential to 
provide vertical support against gravity in the ISM 
\citep{deavillez05,joung09b,hill12,kim13,walch15}, possibly explaining the 
development of the internal turbulent structure inside molecular clouds 
\citep{padoan16}. For example, \citet{kim13} showed using a set of shearing box 
simulations that the stratified medium can be in vertical dynamical equilibrium if 
SN explosions are placed in dense regions ($\nH\ga100\,\cmq$).  The authors 
argue that the turbulent motions that develop due to SNe provide a factor of a few 
larger pressure than the thermal support. In this section, we study the effects of 
different feedback processes on the vertical structure of the ISM.

Following \citet{kim13}, we compute the scale-height as 
$H=\sqrt{\int \rho z^2  dV/ \int \rho dV}$, where the integration is done over the gas 
within the scale-length of the gaseous disk in the $x-y$ plane and $|z|\le 250 \,{\rm pc}$. 
The turbulent and thermal pressure within the scale-length of the disk at the 
mid plane ($|z| \le H$)  is then calculated as 
\begin{align}
P_{\rm th} &= \int P\Theta(\nH<n_{\rm GBC}) dV / \int \Theta(\nH<n_{\rm GBC}) dV, \\
P_{\rm turb} &= \int \rho v_z^2\Theta(\nH<n_{\rm GBC}) dV / \int \Theta(\nH<n_{\rm GBC}) dV,
\end{align}
where $n_{\rm GBC}\approx50 \,\cmq$ is the density of a gravitationally bound cloud 
above which gas is dense enough to self-gravitate, and $\Theta$ is the Heaviside step function.

\begin{figure*}
   \centering
  \raisebox{0.05\height}{\includegraphics[width=5.5cm]{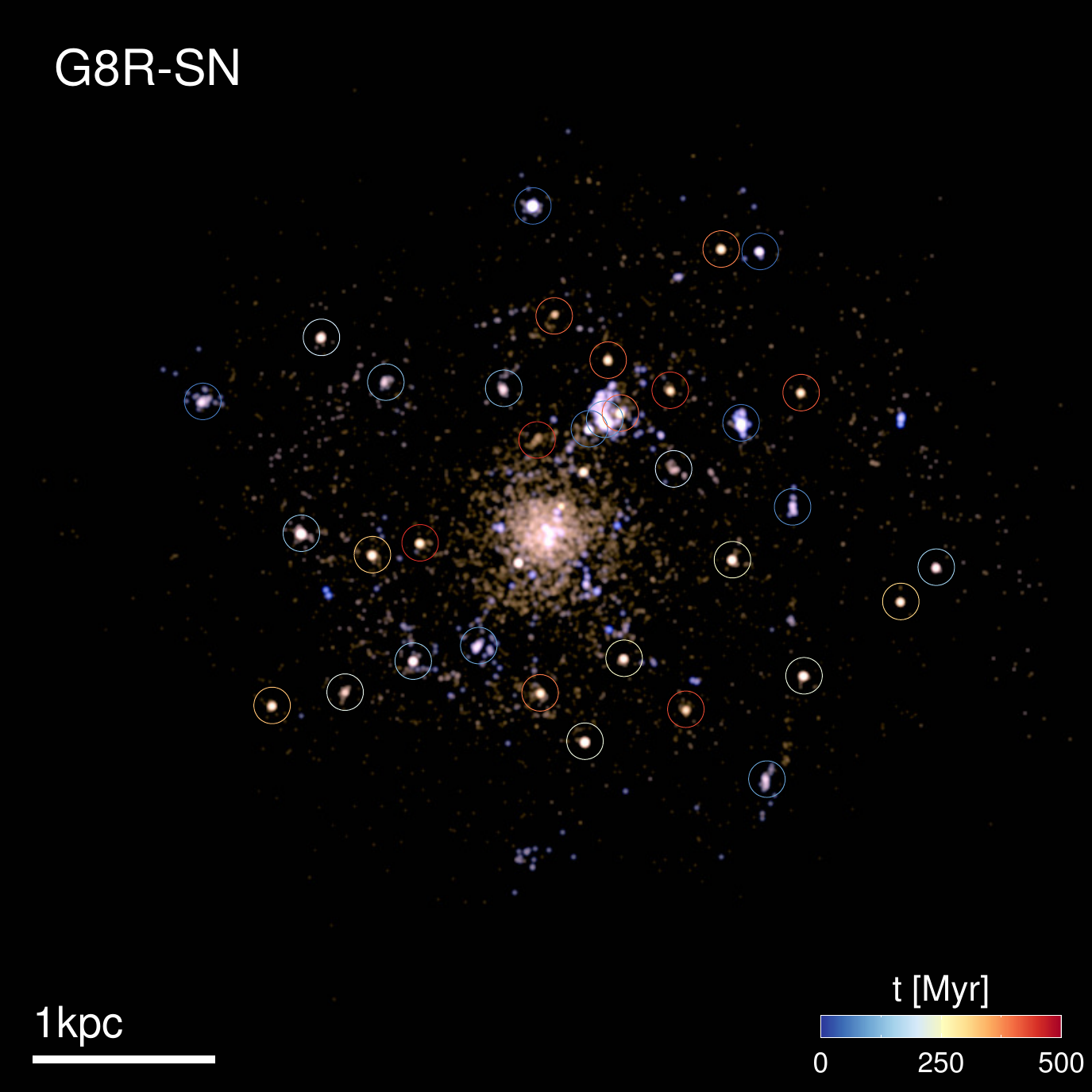}
     \includegraphics[width=5.5cm]{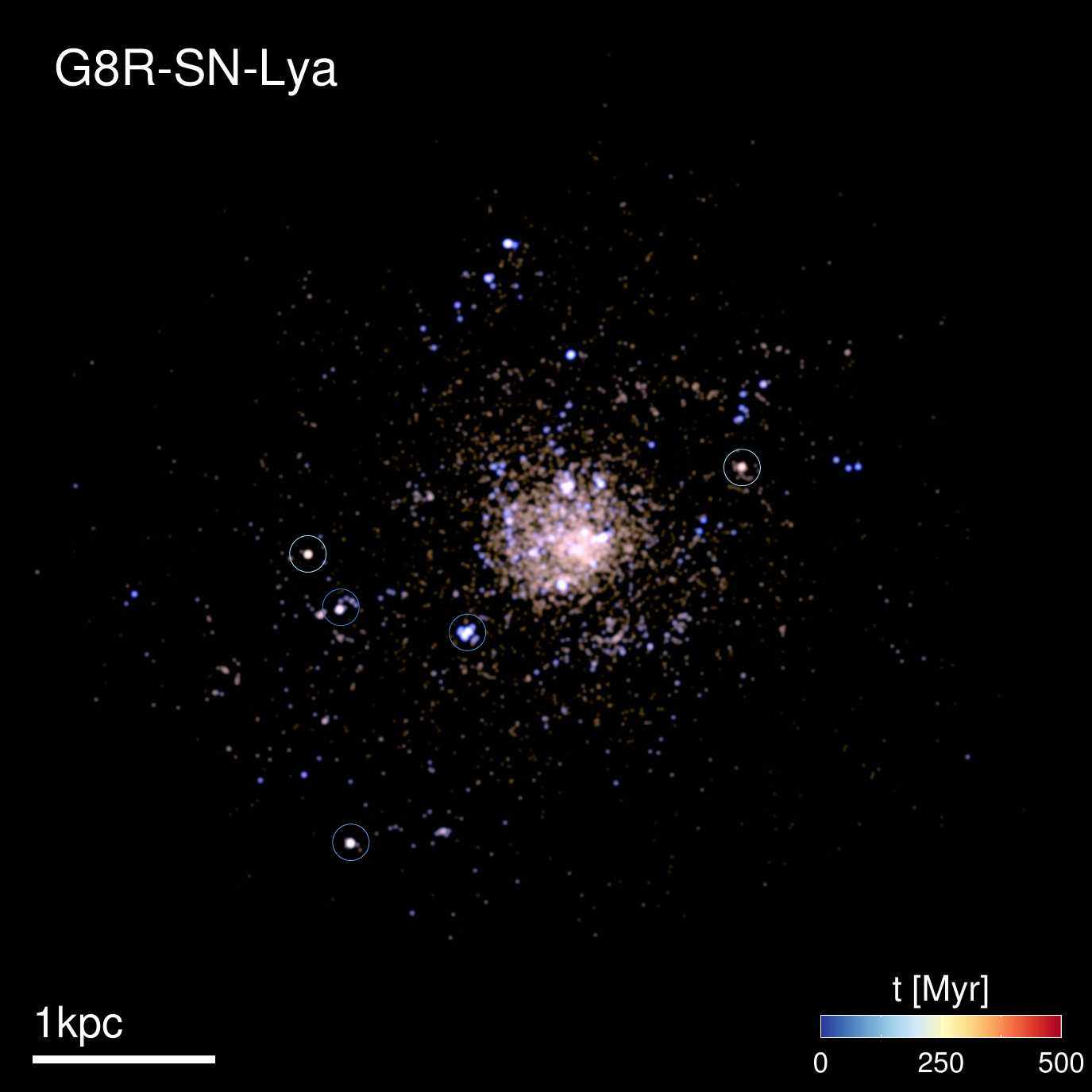}}
  \includegraphics[width=6cm]{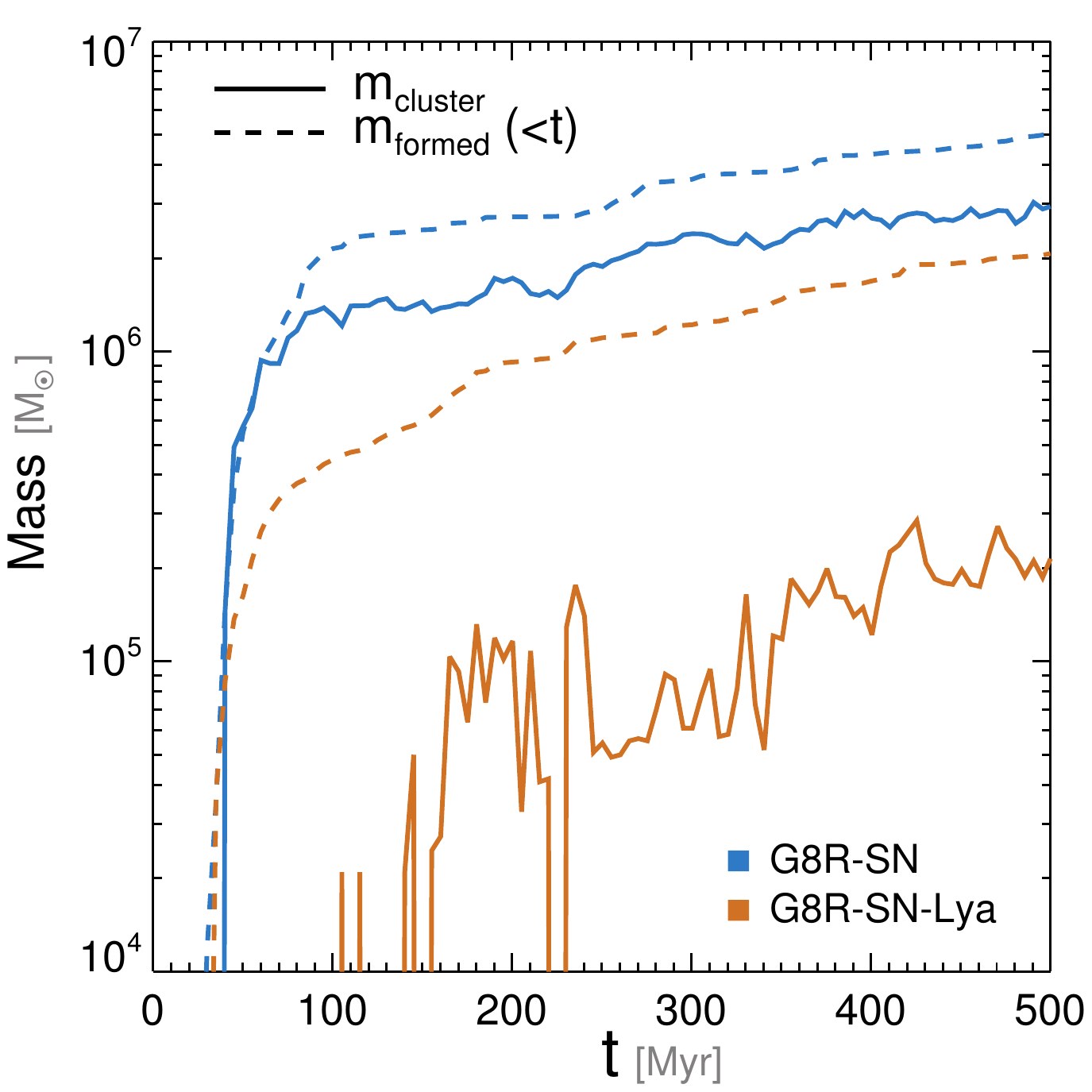}
   \caption{ Star clusters in the run without and with \Lya\ pressure. The composite 
   image is made using the GALEX $NUV$, SDSS $g$, and $i$ bands.  The circles 
   denote the position of star clusters identified using the hierarchical clustering 
   algorithm (HDBscan). The mean age of each star cluster is shown using 
   different colour-codings, as indicated in the legend. The rightmost panel presents 
   the mass enclosed within star clusters outside $0.5\,{\rm kpc}$ as a function 
   of time (soild lines) and the integrated stellar mass formed at $r\ge0.5\,{\rm kpc}$ 
   (dashed lines). Note that the majority ($\sim90\%$) of star particles do not belong 
   to any star clusters in the \texttt{G8R-SN-Lya} run, whereas $\sim50\%$ of the 
   mass is locked up in star clusters in  \texttt{G8R-SN}.
 }
   \label{fig:clusterimg}
\end{figure*}

Figure~\ref{fig:ISM} shows that the disk becomes thicker by adding more energy
from various feedback sources. The disk scale-height is barely resolved in the run 
without feedback ($H\sim20-30\,{\rm pc}$), as the thermal and turbulent pressure are
very low ($\left<P/k_B\right>\sim100\,{\rm cm^{-3}\,K}$). The inclusion of 
photoionization heating increases the thermal pressure substantially 
($\left<P_{\rm th}/k_B\right>\sim10^{4.1}\,{\rm cm^{-3}\,K}$), resulting in 
a thicker disk ($\left<H\right>\approx51\,{\rm pc}$) \citep[see also][]{rosdahl15b}. 
The turbulent pressure is also increased 
($\left<P_{\rm turb}/k_B\right>\sim10^{3.7}\,{\rm cm^{-3}\,K}$),
although it is smaller than $P_{\rm th}$ by a factor of two.
The turbulent pressure becomes stronger (by $\sim30\%$) than the thermal pressure 
when SNe inject the kinetic energy into the ISM (\texttt{G8R-SN}), leading to 
$\left<H\right>\approx64\,{\rm pc}$. \Lya\ feedback further thickens the disk 
($\left<H\right>\approx73\,{\rm pc}$), as both thermal and turbulent pressures 
are enhanced. Note that the mass-weighted average density of the diffuse 
component in the mid plane is roughly a factor of four larger 
($\left<n_{\rm H,0}\right>\approx 8\,\cmq$) in the \texttt{G8R-SN-Lya} model, 
compared to that in \texttt{G8R-SN} ($\left<n_{\rm H,0}\right>\approx 2\,\cmq$), 
due to the less bursty star formation. The run with \Lya\ only  (\texttt{G8R-Lya}) 
exhibits the highest pressure, as the local radiation field is intense owing to the 
less efficient regulation of star formation.

The top left panel of Figure~\ref{fig:ISM} suggests that our simulated galaxies 
with SN feedback are in approximate pressure equilibrium. The thermal and 
turbulent pressure remain around $P/k_B \sim 10^4 - 10^5 \, {\rm cm^{-3}\, {\rm K}}$,
which is similar to the pressure required to counter-balance vertical weight of 
the ISM ($\approx 10^{4}\, {\rm cm^{-3}\, {\rm K}} (\Sigma / 10 \,\msun\,{\rm pc}^{-2}) 
\sqrt{\rho_{\rm sd}/0.1\,\msun \, {\rm pc}^{-3}}$), where
$\Sigma\approx 20-30\,\msun\,{\rm pc}^{-2}$  is the gas surface density, 
and $\rho_{\rm sd}\approx 0.1-0.2\,\msun\,{\rm pc}^{-3}$ is the mid-plane density 
due to stars and dark matter \citep{ostriker10,kim13}. The two models 
(\texttt{G8R-SN} and \texttt{G8R-SN-Lya}) are also in good agreement with the fitting 
results from the local shearing box simulations \citep{kim13} where the ISM 
is shown to be in vertical equilibrium. The main difference is that turbulent support 
in our simulations does not appear to contribute substantially to the total pressure 
as in \citet{kim13}. For example, their gas-rich ($\Sigma=20\,\msun\,{\rm pc}^{-2}$) 
models (QA20 series), which is comparable to our conditions 
($\left<\Sigma\right>\approx33\,\msun\,{\rm pc}^{-2}$), exhibit a factor of 2--3 larger 
turbulent support than thermal pressure, but we found that the turbulent pressure is
stronger than thermal pressure only by $\sim50\%$. We attribute this to the fact that 
photo-ionization heating provides a extra thermal pressure to the diffuse gas in the 
mid plane. Indeed, when we do not consider any type of radiation feedback 
(\texttt{G8SN}), the turbulent support is found to be roughly twice as strong as the 
thermal support \citep[see also][]{kim13}. As a result, the disk thickness in this run 
($\left<H\right>\sim48\,{\rm pc}$) is smaller than in the run with radiation.

\subsection{Formation of massive star clusters}

Given the large multiplication factor of \Lya\ photons in the metal-poor environment, 
the relative difference in stellar mass between \texttt{G8R-SN} and 
\texttt{G8R-SN-Lya} may not appear very remarkable. However, 
Figure~\ref{fig:clusterimg} demonstrates that \Lya\ feedback plays a crucial role 
in shaping the formation of star clusters. To quantitatively measure this, we use a hierarchical clustering algorithm based on the mutual 
reachability distance, HDBscan\footnote{http://hdbscan.readthedocs.io/en/latest/index.html} \citep{Campello2013}. 
The algorithm generates the member list of each cluster without assuming any particular
structure. We determine the centre by iteratively computing the centre of mass 
within a fixed 100 pc radius.  A cluster with fewer than 20 particles or a structure 
that is not centrally well concentrated 
(i.e., $M_{\rm star}^{\rm r<50pc} / M_{\rm star}^{\rm r<100pc} > 0.5$) is removed 
from the cluster list. 

The number of star clusters present in the final snapshot ($t=500\,{\rm Myr}$) is 34
in \texttt{G8R-SN}, whereas only 5 clusters are identified in the fiducial run. 
Note that we ignore any clusters located in the inner region ($r<0.5\, {\rm kpc}$), 
because we are mainly interested in the survivability of star clusters against 
internal feedback processes. 
The final mass contained in the star clusters outside 0.5 kpc in the \texttt{G8R-SN} 
and  \texttt{G8R-SN-Lya} run is $2.3\times10^6\,\msun$ (35.3\% of the 
total stellar mass) and $1.7\times10^5\,\msun$ (4.4\%), respectively. 
Individual star clusters are also less massive in the run with \Lya\ feedback. The typical mass of a star cluster found in the 
simulations is $5.1\times10^4\,\msun$ and $2.6\times10^4\,\msun$, respectively.

Figure~\ref{fig:clusterimg} also shows that \Lya\ pressure makes massive star
clusters more difficult to  form and survive. 
The rightmost panel displays the integrated mass of stars formed 
outside 0.5 kpc (dashed lines) and the mass enclosed within clusters as a function 
of time. Here we combine the initial mass of star particles for the enclosed mass 
instead of actual particle mass at time $t$ in order to directly compare with the 
stellar mass formed. There are considerably more old star clusters 
in the \texttt{G8R-SN} run, while only young star clusters are left in 
\texttt{G8R-SN-Lya} (see the colour of the circles). It is also evident from this plot that a large fraction of stars 
($\sim90\%$) do not belong to any star cluster in the \texttt{G8R-SN-Lya} run. 
This indicates that \Lya\ pressure regulates the efficiency for an individual star formation event from the early stage of cluster formation.
It is also likely that small clusters are formed and then dispersed as early feedback 
processes blow away the self-gravitating gas through non-adiabatic expansion 
\citep{pontzen12,teyssier13}.
The rapid variation of the mass enclosed within star clusters corroborates that  
star clusters are short-lived in the presence of strong early feedback. 
On the contrary, \texttt{G8R-SN} appears to convert the gas into stars more efficiently,
turning them into a self-gravitating object. As a result, a large fraction of star clusters 
($\sim50\%$) survive SN explosions.

Some authors claim that globular clusters (GCs) may form in the gas-rich 
galactic disc \citep[e.g.][]{kravtsov05}, and one may wonder how the star 
clusters identified from our simulations compare with the properties of GCs. 
Since the half-mass radii of the clusters are barely resolved in our simulations 
($\approx14\,{\rm pc}$), we focus on the age and mass.
The simulated star clusters are less massive than the GCs ($\sim10^5\,\msun$) 
by a factor of a few, and show significant internal age dispersions, 6~Myr 
(\texttt{R-SN-Lya}) or 24~Myr (\texttt{R-SN}), respectively. The maximum mass 
of the clusters ($3.0\times10^5\,\msun$ and $8.9\times10^4\,\msun$ in the run 
without and with \Lya\ pressure) is also smaller than what is typically derived 
in observations ($\sim10^6\,\msun$). More importantly, the short lifetimes of the 
star clusters in the \texttt{G8R-SN-Lya} run do not seem to favour the scenario
in which GCs form inside a metal-poor, gas-rich disk,  and survive to the present day.
We note, however, that the observed specific frequency of the GCs in dwarf galaxies 
comparable to our simulated one ($L_V\sim-14$) ranges from 0 to 5 \citep[e.g.,][]{georgiev10},
and thus the short lifetimes do not necessarily rule out the general possibility of forming a GC 
in gas-rich, massive spiral galaxies.

Our experiments demonstrate that \Lya\ pressure may be a critical mechanism 
by which potential globular cluster candidates with low metallicities are disrupted. 
Even if the mechanism cannot disrupt a GC, it may help to suppress the internal 
age and metallicity dispersions of the clusters by truncating star formation early on 
\citep[e.g.][]{kimm16}. In this regard, future simulations aiming to understand
the detailed formation histories of metal-poor GCs may need to include 
\Lya\ feedback along with other radiation feedback processes.

\begin{figure}
   \centering
     \includegraphics[width=8.5cm]{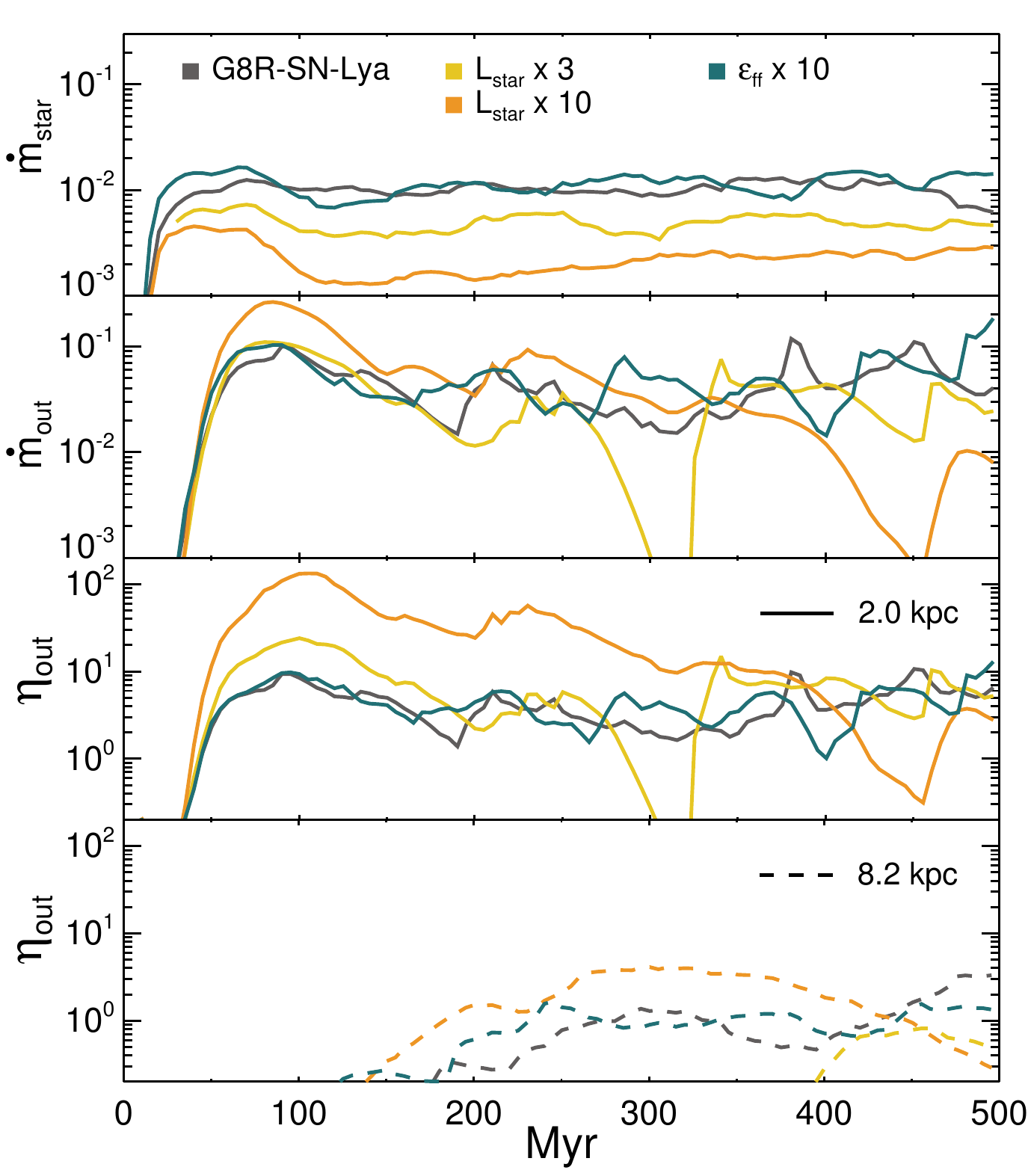}
   \caption{  
   Effects of the different radiation field strength and star formation efficiency. The models in which 
   the luminosity of stars is boosted by a factor of 3 (\texttt{G8R-SN-Lya-f3})
   or 10 (\texttt{G8R-SN-Lya-f10}) are shown as yellow and red colours, respectively.
   The blue lines show the case in which $\epsilon_{\rm ff}$ is boosted by an order 
   of magnitude (\texttt{G8R-SN-Lya-s10}). From top to bottom, each panel displays
   the star formation rates, outflowing rates, mass-loading factors ($\eta_{\rm out}$) 
   measured at $|z|=2\,{\rm kpc}$, and $\eta_{\rm out}$ measured at 
   $0.2\,R_{\rm vir}=8.2\,{\rm kpc}$. Note that $\eta_{\rm out}$ at $0.2\,R_{\rm vir}$ 
   is smaller than $\sim10$ even with extreme feedback  (\texttt{G8R-SN-Lya-f10}).
   }
   \label{fig:extreme}
\end{figure}

\begin{figure}
   \centering
     \includegraphics[width=8.5cm]{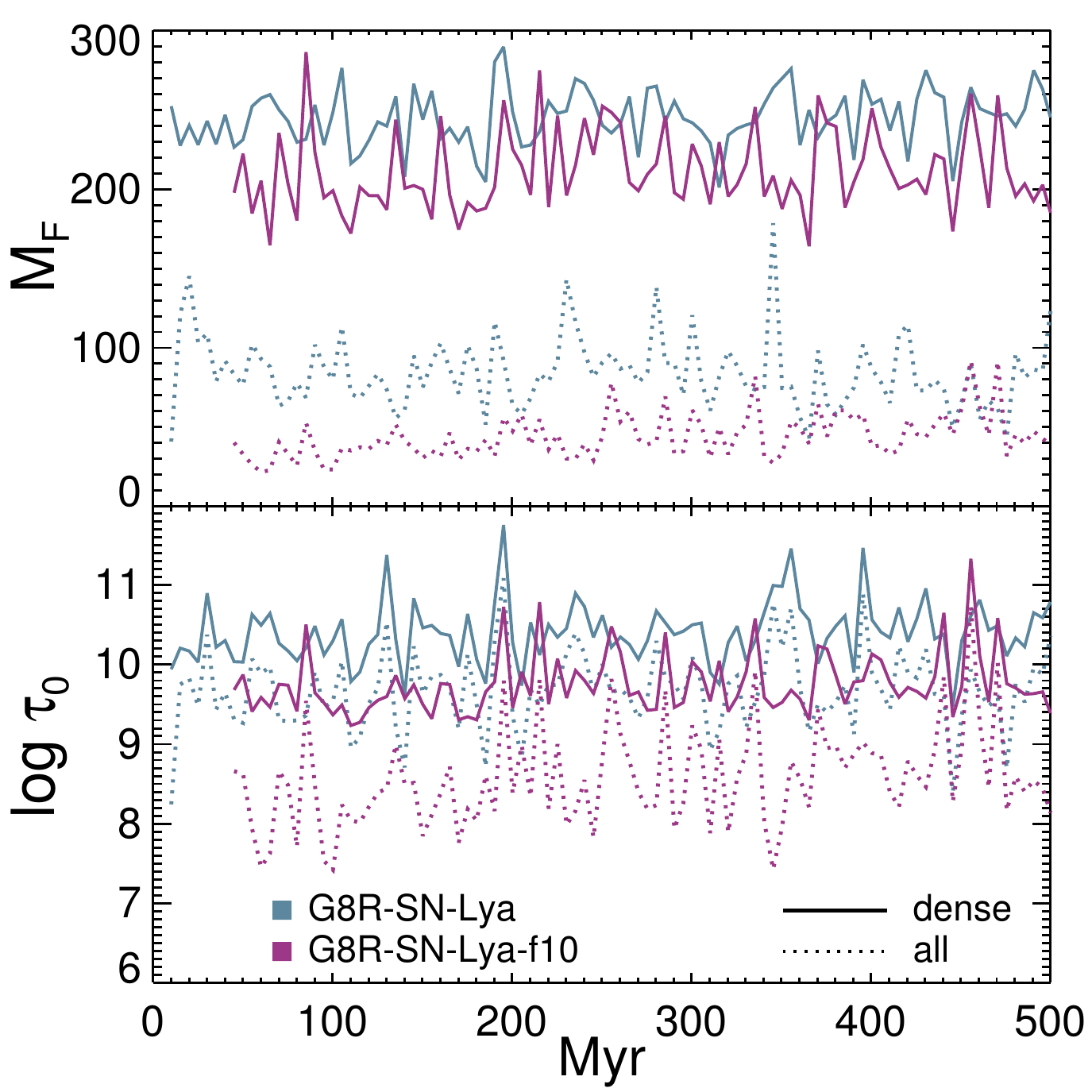}
   \caption{  
The photon number ($N_{\rm Lya}$)-weighted average of the multiplication factors 
(top panel) and optical depths to \Lya\ photons at the line centre (bottom panel). 
The solid lines show $M_F$ and $\tau_0$ of the dense regions ($\nH\ge100\,\cmq$),
whereas the dotted lines show the average for the entire gas. The blue and violet 
colours indicate the runs with the normal and extreme radiation field.
} 
   \label{fig:tau}
\end{figure}

\section{Discussion}

\subsection{Effects of early stellar feedback}

We have demonstrated in the previous section that the inclusion of \Lya\ pressure 
disrupts gas clouds more efficiently than the run without it. 
Nevertheless, the mass-loading factors are decreased, compared to the run without 
\Lya\ feedback, as star formation becomes less bursty and SNe do not explode 
in a collective manner. This raises the question as to whether or not early stellar feedback
can help drive strong winds that carry a large amount of gas in low-mass 
haloes \citep[e.g.][]{muratov15}. To investigate this, we run two additional simulations
by artificially increasing the luminosity of individual stars by a factor of 3 
(\texttt{G8R-SN-Lya-f3}) or 10 (\texttt{G8R-SN-Lya-f10}). By doing so, we enhance 
not only the pressure from multi-scattered \Lya\ photons but also photo-ionization 
heating and radiation pressure from UV and IR, although the most significant impact
still originates from \Lya\ feedback.
Note that the inclusion of higher mass stars in a simple stellar population 
up to $M_{\rm max}=300\,\msun$, as opposed to $M_{\rm max}=100\,\msun$, 
can easily boost the number of ionizing photons
by a factor of $\approx 2$. In this regard, \texttt{G8R-SN-Lya-f3} may be taken as 
the optimistic but still plausible model. However, \texttt{G8R-SN-Lya-f10} is certainly 
an extreme case, which may be used to learn about the effects of early feedback.

Figure~\ref{fig:extreme} shows that star formation rates are directly affected 
by the strength of radiation field from stars in the presence of  \Lya\ pressure. 
The total amount of stars formed during $500\,{\rm Myr}$ is reduced by a factor 
of $\approx$2 or 4 in the runs with three times or ten times more photons 
($2.4\times10^6\,\msun$ and $1.2\times10^6\,\msun$ for the \texttt{f3} and 
\texttt{f10} runs, respectively). Note that the suppression of star formation
is not simply proportional to the boost factor. This is partly because \Lya\ photons 
cannot reside in an extremely optically thick region for a long time if star-forming 
clouds are irradiated by intense radiation. Figure~\ref{fig:tau} substantiates this by 
showing that the $N_{\rm Lya}$-weighted mean multiplication factor is systematically 
reduced from $\left<M_F\right>=244$ to $212$ in dense regions ($\nH\ge100\,\cmq$). 
The difference is more noticeable in the less dense environments 
($\nH<100\,\cmq$, $\left<M_F\right>=91.2$ vs $47.6$), meaning that the ISM 
becomes more diffuse under strong radiation feedback. Figure~\ref{fig:tau} 
(bottom panel) indeed shows that the optical depth to \Lya\ photons
at the line centre is decreased by an order of magnitude around young stars.

Outflow rates are adjusted in such a way that the mass loading factors measured 
at $|z|=2\,{\rm kpc}$ are around 2--10 between $300\le t \le 500\,{\rm Myr}$. 
The extreme model displays $\eta_{\rm out}$ up to $\sim100$ in the initial phase during 
which the mean density of the outflow is temporarily augmented by a factor of two 
and the covering fraction of the outflows is increased by a factor of five.  
However, the velocity ($10-30\, \kms$) and temperature ($T\sim10^5\,{\rm K}$) of the 
outflows are largely unchanged despite the significant decrease in star formation rates. 
The condition for such a gas-rich environment may be met during the starburst phase 
(e.g. mergers). Even when radiation feedback is artificially boosted by an 
order of magnitude, the mass loading factor measured at $0.2\,{\rm \rvir}$,  especially at late times, turns out 
to be remarkably smaller than previous studies which successfully reproduce a 
variety of observations \citep[see also][]{rosdahl15b}. For example, \citet{muratov15} obtained 
$\eta_{\rm out}\sim40-100$ with the median velocity of 
$v_{\rm wind}\sim40-80\,\kms$ in halos of mass adopted in this study.
The momentum-driven wind model by \citet{oppenheimer08} used a more similar mass 
loading ($\eta_{\rm out}\approx 5$ if we assume $\sigma\approx V_c/\sqrt{2}$),
although the wind velocity at the time of injection is slightly higher than our predictions.

One of the main differences from the simulations used in \citet{muratov15} is the 
way star formation is modelled. Although the idea behind the use of local gravity is 
very similar, the efficiency for star formation per free-fall time is set to 100\% in the 
FIRE simulation \citep{hopkins14}, while the typical efficiency is 16\% in the 
\texttt{G8R-SN-Lya} run. Although $\epsilon_{\rm ff}=16.4\%$ is still higher than 
what we find on galactic scales in the local Universe \citep{kennicutt98,evans14}, 
the finite resolution of our simulations may require even a higher efficiency to take the
best advantage of clustered SN explosions. Motivated by this, we 
also test a model in which $\epsilon_{\rm ff}$ is boosted by an order of magnitude 
(\texttt{G8R-SN-Lya-s10}). Because adding extra pressure in star-forming regions 
from radiation feedback usually reduces $\epsilon_{\rm ff}$, we find that the 
resulting $\epsilon_{\rm ff}=1.37$ is slightly smaller than 
1.64 ($10\times\epsilon_{\rm ff}$ from \texttt{G8R-SN-Lya}), but still significantly 
higher than the fiducial run. Figure~\ref{fig:extreme} shows 
that the models yield very similar results not only in terms of star formation rates 
but also in outflow rates. We find that the properties of the outflows are also 
indistinguishable between the enhanced $\epsilon_{\rm ff}$ model and the fiducial model, 
although the escape fractions of LyC photons are increased (see the next section). 
This is again due to the fact that \Lya\ pressure controls the dynamics of 
star-forming clouds and suppresses very bursty star formation episodes. 

To summarise, these extreme models demonstrate that strong 
early feedback does not necessarily enhance outflow rates, but mostly acts to 
stabilise the star formation histories of the galaxy, reducing the mass-loading factor.
This calls into question how gas can be entrained at truly enormous mass loading factors ($\eta_{\rm out}\sim 50$--$100$).
Unfortunately, there are only a 
handful of local analogues of high-$z$ dwarf galaxies. Furthermore, making a 
direct comparison to 
these observations is not straightforward. \citet{heckman15} shows that dwarf 
galaxies with circular velocities of $v_{\rm cir}\sim40\,\kms$ tend to show 
intermediate mass loadings of $\dot{M}_{\rm out}/\dot{M}_{\star} \sim 6-8$, 
but these systems appear to have larger stellar masses and star formation rates ($\dot{M}_\star\sim0.1\,\msun\,{\rm yr^{-1}}$).
The most relevant sample from \citet{heckman15} is I Zw 18, which has a very similar 
metallicity and star-formation rate as our simulated galaxies, but only an upper limit 
on the mass loading of $\dot{M}_{\rm out}/\dot{M}_{\star} < 60$ is placed due to 
the ambiguity of outflow velocities ($v_{\rm out}$). Moreover, it should be noted that 
these estimates are computed by counting all the mass along the line of sight 
($N_{\rm out} \left< m\right>$) and assuming that outflows are launched 
at twice the starburst radius ($r_{\rm out}\sim2\,r_*$), 
i.e. $\dot{m}_{\rm out} = 4 \pi N_{\rm out} \left<m\right> v_{\rm out} 2\,r_*$, 
where $N_{\rm out}$ is the column density of the outflow, and
$\left<m\right>$ is the mean mass per particle. Since the outflow rates can be written
as $\dot{m}_{\rm out} (r) = 4\pi r^2 \rho v_r$, the above equation implicitly assumes that 
the surface density is $\Sigma=N_{\rm out}\left<m\right> \sim r_{\rm out}\rho(r_{\rm out})$.
For outflowing gas that follows an isothermal profile\footnote{
\citet{muratov15} shows that the outflowing mass contained in equally binned shells in dwarf-sized halos
is more or less constant, indicative of $\rho\propto r^{-2}$.}
with a truncation radius at $a$, $\rho = A / (r^2 + a^2)$, 
one can show that the integrated surface density from the centre to a radius $R$ is 
$\Sigma(\le r)= \frac{A}{a} \tan^{-1} \left(\frac{r}{a}\right)$, 
where $A=\frac{M_{\rm out}}{4\pi \left(R_{\rm vir} - \tan^{-1}[R_{\rm vir}/a] \right)}$. 
Thus, the integrated surface density, $\Sigma(<R_{\rm vir})\sim N_{\rm out}\left<m\right>$,
used in the observation to derive outflow rates is likely to be larger than 
$r_{\rm out}\rho(r_{\rm out})$ by 3--4 times for $r_* \le a \le 3 r_*$,
and any discrepancy (or agreement) between observations and simulations 
should be taken with caution. Of course, if the outflows are not mass conserving \citep[e.g.][]{leroy15}, the discrepancy would be smaller. Recent work by \citet{chisholm17} takes 
into account this possibility along with the photoionization on ultraviolet spectra
from the Cosmic Origin Spectrograph on the Hubble Space Telescope, and 
conclude that the {\em maximum} mass loading factors of the 
systems (SBS 1415+437 and I Zw 18) that share a similar star formation rate
($0.02\,\msun\,{\rm yr^{-1}}$) are $19\pm 17$ and $11\pm8.0$, respectively.
Intriguingly, the mass-loading from our fiducial model does not seem in stark 
contradiction to these estimates. However, in order to draw a firm conclusion,
it will be necessary to generate mock absorption spectra and re-derive the 
mass loading based on the same methodology used in observations.

\begin{figure*}
   \centering
     \includegraphics[width=8.5cm]{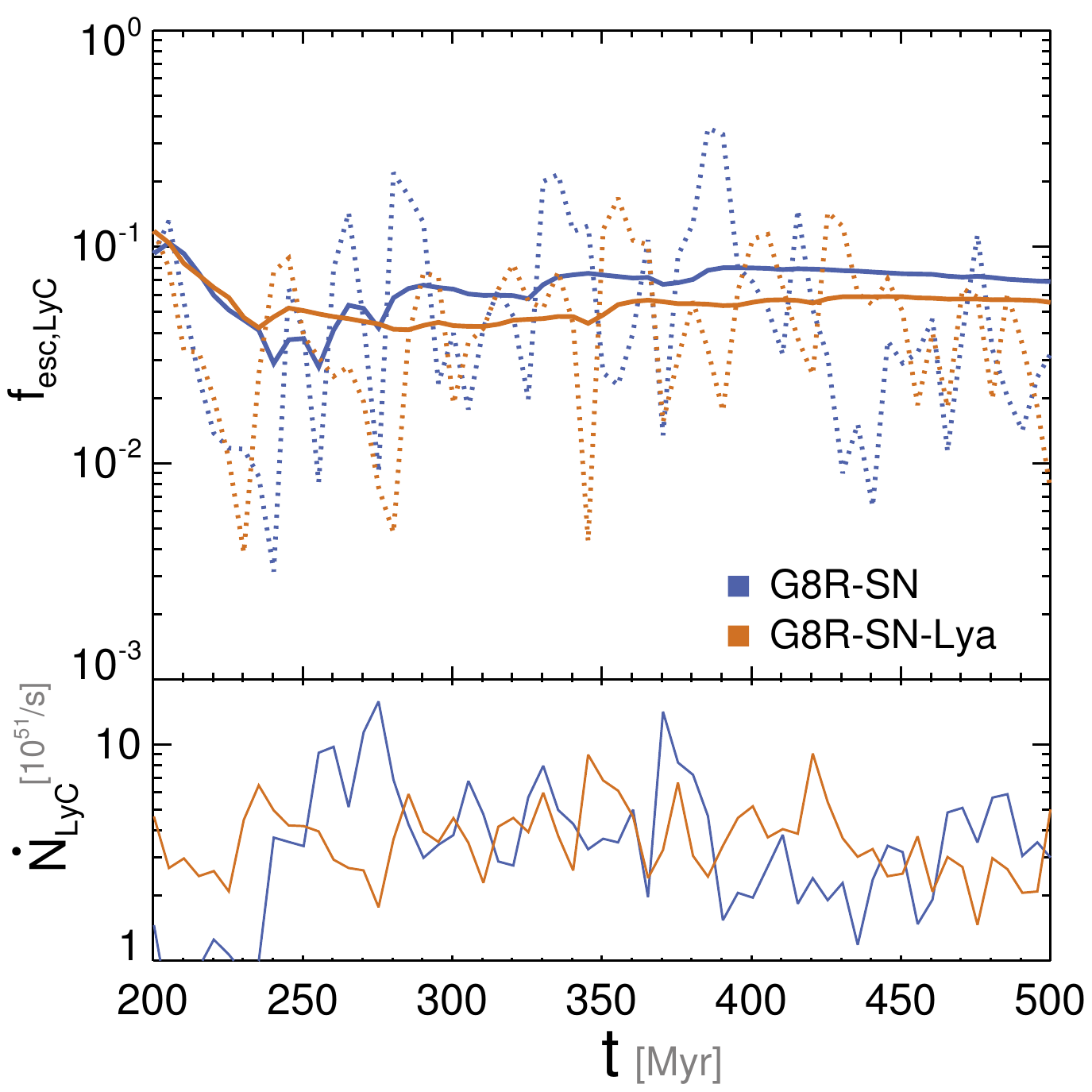}
       \includegraphics[width=8.5cm]{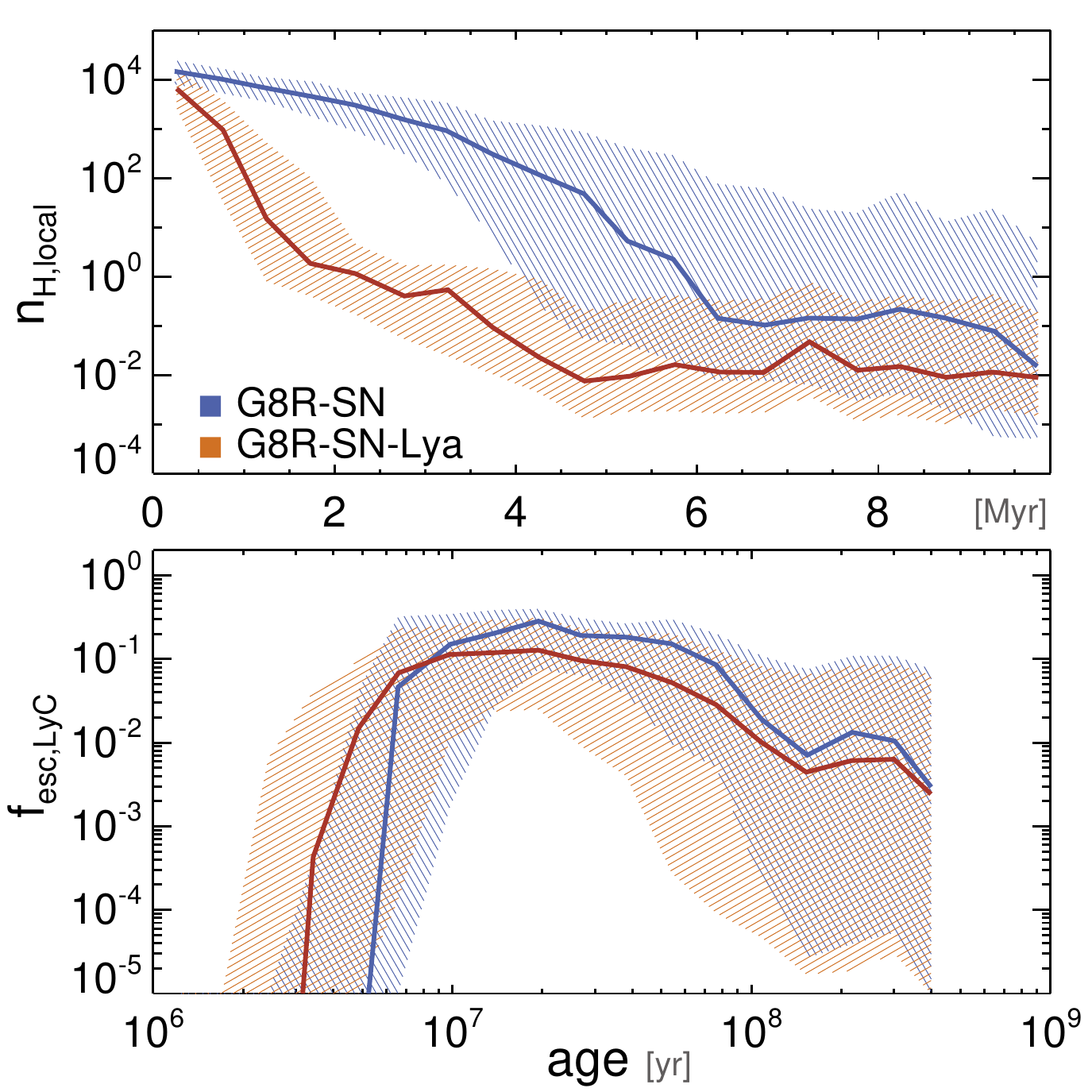}
   \caption{  
{\em Left}: Escape fractions and production rates of LyC photons in two different runs. 
The dotted and solid lines show the instantaneous and photon-number averaged 
$f_{\rm esc,LyC}$. {\em Right}: the local density in which star particles with 
different ages are located (top) and the median instantaneous $f_{\rm esc,LyC}$ as 
a function of the age of star particles. The shaded regions mark the interquartile range of the distributions. Although \Lya\ feedback lowers the density around young 
stellar populations, the average escape fractions turn out to be smaller than those 
from the \texttt{G8R-SN} run (see the text). 
}
   \label{fig:fesc}
\end{figure*}

\subsection{Possible implications for reionisation}

Studies show that the disruption of star-forming clouds is crucial for ionizing 
photons to escape from their host dark matter haloes and reionize the Universe 
\citep{wise09,dale12,kimm14,wise14,paardekooper15,kimm17,trebitsch17}.
Given that \Lya\ feedback accelerates the disruption process,
it is necessary to check whether or not the mechanism may change the previous 
picture of reionization. To investigate this in detail, we measure escape 
fractions by computing the escaping probability of LyC photons on the virial sphere
of the dark matter halo as follows. We use the {\sc healpix} algorithm \citep{gorski05} 
to generate 2048 photons per star particle, which carries information about a spectral energy distribution
as a function of wavelength for a given age, mass, and metallicity. The photons are attenuated by neutral hydrogen, 
as $f_{\rm abs} (\nu)  = f_{\rm int} (\nu) \exp{\left[-\tau_{\rm HI} (\nu)\right]}$, 
where  $\tau_{\rm HI}$ ($=N_{\rm HI} \sigma_{\rm HI}$) is the optical depth and 
$\sigma_{\rm HI}$ is the hydrogen absorption cross section \citep{osterbrock06}.
We also take into account the small magellanic cloud-type dust \citep{weingartner01},
but its contribution turns out to be negligible even with the high dust-to-metal ratio 
of 0.4.

The left panel of Figure~\ref{fig:fesc} presents the instantaneous (dotted lines) and photon-number 
weighted average escape fractions (solid lines)  of the two most interesting runs 
(\texttt{G8R-SN} and \texttt{G8R-SN-Lya}) at $200<t<500~{\rm Myr}$. 
We do not include the initial phase of gas collapse, because the ISM structures 
are likely to be affected by the initial condition, although including the results 
does not change any of the conclusions that follow. The figure also includes the photon production rates 
in the bottom left panel, and one can see that the escape fractions tend to be higher after strong starburst,
which is consistent with  previous findings from cosmological RHD simulations \citep{kimm14,wise14,kimm17,trebitsch17}.

Interestingly, we find that a slightly smaller fraction of the ionizing photons escapes from the dark 
matter halo in the run with \Lya\ feedback. The photon-number weighted average 
escape fractions at $200<t<500~{\rm Myr}$ are measured to be $\left<f_{\rm esc,LyC}\right>=5.5\%$ and 
$\left<f_{\rm esc,LyC}\right>=6.9\%$ for the runs with and without \Lya\ feedback,
 respectively. This is somewhat counter-intuitive, given that the inclusion of early 
 feedback is expected to lower the local gas density surrounding stars. The right panel of 
Figure~\ref{fig:fesc} indeed demonstrates that the gas density of the cells in which star 
particles older than $\sim 2\,{\rm Myr}$ live is $\nH\la1\,\cmq$, which is much 
lower than the \texttt{G8R-SN} case. These cells are sufficiently diffuse to become ionized
by a single star particle of mass $910\,\msun$, and we also confirm that  
the ionization fraction of the cells hosting stars older than $\sim2~{\rm Myr}$ in the \texttt{G8R-SN-Lya} 
run is close to unity.

\begin{figure}
   \centering
     \includegraphics[width=8.cm]{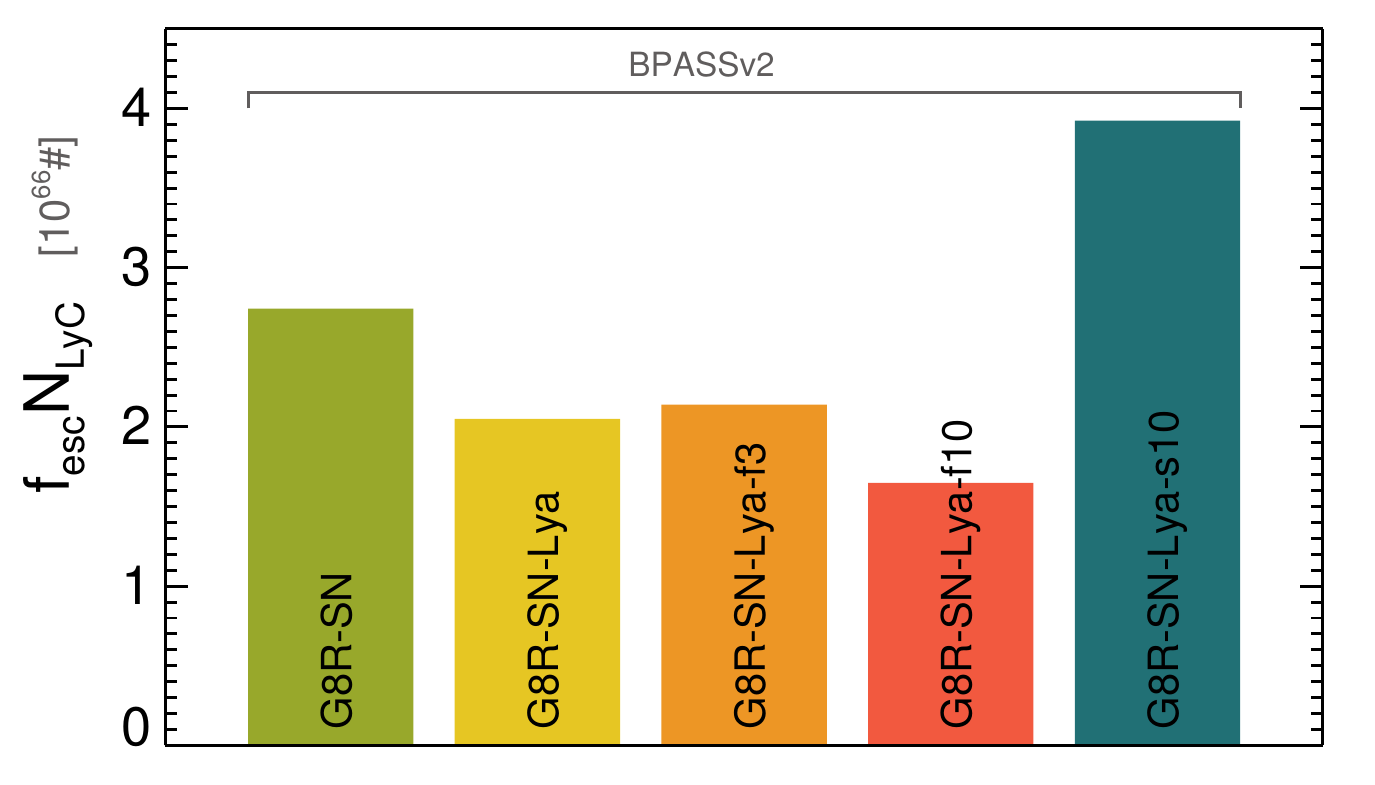}
   \caption{  
The integrated number of escaping LyC photons in the simulations with various 
feedback strength between $200\le t \le 500\,{\rm Myr}$. It can be seen that the 
number of escaping photons are not very sensitive to the strength of radiation 
feedback, but it depends more on the star formation model. }
   \label{fig:nph_fesc}
\end{figure}

The lower escape fractions in the  \texttt{G8R-SN-Lya} run can be understood 
in terms of the efficiency of creating channels transparent to LyC photons.
\citet{kimm17} show that the escape fractions can be as high as 
$f_{\rm esc,LyC}\sim 40\%$ if a large number of stars are formed instantaneously 
and the birth clouds are disrupted by photo-ionization heating. However, if the cloud 
is massive enough to delay and confine the propagation of ionizing radiation within 
the galactic ISM, the escape fractions are unlikely to be high. 
Figure~\ref{fig:fesc} (right panel) supports this claim by showing that 
$f_{\rm esc,LyC}$ of stars younger than $t\sim10\,{\rm Myr}$ is very low\footnote{
Note that since the outer region of the simulated halo is hot, 
all of the absorption takes place within the galaxy, and any photons 
that leave the galaxy can easily reach the virial radius.}.
On the contrary, stars older than $t\sim10-100\,{\rm Myr}$ exhibit a higher 
$f_{\rm esc,LyC}$ of $\sim10-30\%$. The fact that \texttt{G8R-SN} displays a higher 
average $f_{\rm esc,LyC}$ in these stars suggests that 
more coherent Type II SNe from bursty star formation are better at creating HII holes 
than \Lya\ pressure, as the latter is a gentle process that does not drive extended 
outflows. Stars older than 100 Myr exhibit lower escape fractions again, 
because they cannot emit a large number of LyC photons and the resulting
Stromgren sphere is very small.

However, when radiation pressure by \Lya\ photons dominates over the effects due 
to SNe, the escape fractions of ionizing radiation are increased significantly.
For the runs employing a stronger radiation field (\texttt{G8R-SN-Lya-f3}), 
$11.6\%$ of ionizing photons leak out from the dark matter halo. 
The escape fraction is further increased to 18.5\% in the \texttt{G8R-SN-Lya-f10} run.

The balance between enhanced escape fractions and reduced star formation 
implies that the reionization history of the Universe may not be significantly affected  
by the presence of \Lya\ feedback. Figure~\ref{fig:nph_fesc} shows that 
the integrated number of escaping LyC photons in the plausible models 
(\texttt{R-SN}, \texttt{R-SN-Lya}, \texttt{R-SN-Lya-f3}) is reasonably similar.
Even the extreme feedback model produces  a mere 20\% fewer escaping photons than 
those from \texttt{R-SN-Lya}. In a slightly different context,
\citet{kimm14} also report that a similar number of LyC photons escapes from dark 
matter halos irrespective of the presence of runaway stars,
because runaway stars suppress star formation but enhance escape fractions.

However, we find that the predicted escape fractions rely on the choice of the star 
formation model. In the \texttt{G8R-SN-Lya-s10} case where star formation is 
designed to be locally more bursty, radiation feedback comes into play earlier than in 
the case of \texttt{G8R-SN-Lya}, lowering the local densities of young stars 
($\la5\,{\rm Myr}$) even further than the fiducial model. As a result, 
$\left<f_{\rm esc,LyC}\right>$ is elevated to $8.8\%$, which is roughly twice
larger than the fiducial case despite the fact that a similar amount of stars 
is formed in both runs. Even though we prefer our star formation
model in which the efficiency per free-fall time is calculated based on the local
virial parameter and turbulence, the uncertainty is still worrisome,
and will be required to test against well resolved, cloud simulations \citep[e.g.][]{dale12,geen16,gavagnin17}.

\subsection{Caveats}

We emphasise that there are several important caveats in the modelling of 
\Lya\ feedback. First, the amount of dust in high-z metal-poor galaxies is highly 
uncertain. We adopt the values derived from the local dwarf galaxies 
\citep{remy-ruyer14},  but these estimates are still uncertain due to small number 
statistics. Nonetheless, there is a clear indication that the relative mass fraction of dust 
with respect to metal mass is much smaller than those of local spiral galaxies \citep{lisenfeld98,engelbracht08,galametz11,fisher13,remy-ruyer14}.
If we assume that the dust-to-metal ratio is only 20 times smaller than the local 
solar neighbourhood (i.e. $f_{\rm d/m}=0.05$), instead of $f_{\rm d/m}=0.005$
we adopt in this study, the maximum multiplication factor will decrease from 
$M_{F,\rm max}=355$ to $M_{F,\rm max}=216$. This will give similar effects 
as decreasing the luminosity of stars by $\approx 40\%$. However, given that the 
number of ionizing photons from a simple stellar population can be augmented 
by a factor of two when the formation of massive stars ($M\ge100\,\msun$) is allowed,
the effects due to \Lya\ scattering are not probably severely over-estimated.

Second, even though the radial momentum from \Lya\ pressure is estimated based on 
a Monte Carlo \Lya\ radiative transfer code ({\sc rascas}), our subgrid model cannot 
account for the long range force due to spatially diffused \Lya\ photons after scattering. 
Such a calculation would require fully coupled radiative transfer, as in \citet{smith17,smith17b}, but 
this is not yet feasible for three dimensional galactic-scale simulations\footnote{We 
have estimated the computational costs assuming that the MCRT step is done 
only in every coarse time steps, and found that including the calculations with 
a small number of photons ($N_{\rm Lya}=1000$) can slow down the simulation
by more than an order of magnitude, compared to a typical RHD run.}. As a first step, we focus 
on the short-range force that primarily affects the evolution of star-forming clouds.
Note that this is a reasonable approximation, as long as the majority of \Lya\ photons
are destroyed by dust inside the GMC. Indeed, as shown in Figure~\ref{fig:tau}, 
the multiplication factors from \texttt{G8R-SN-Lya} are already close to the maximum 
value one can exploit from the metal-poor environments ($M_{F,\rm max}\approx355$). 
If we limit our discussions to denser, star-forming regions ($\nH\ga10^3\,\cmq$), 
the agreement is better ($\left<M_F\right>\sim300$), meaning that most \Lya\ photons
are absorbed within our computational resolution. It is possible that 
we over-estimate the feedback effects around ionization fronts formed in 
low-density environments because 
the light travel time for multi-scattered photons can be longer 
than the simulation time step. However, given that the local photon density 
and the multiplication factors are usually small ($M_F\la 10$) in these 
regions, we expect that photoionization heating governs the local gas dynamics, 
and the uncertainty due to this simplification is likely insignificant.

Another important uncertainty is the lack of internal structures in the 
star-forming regions. Although our maximum resolution (4.6 pc) is 
substantially higher than recent large-scale cosmological simulations \citep[e.g.][]{dubois14,vogelsberger14,schaye15,dave17} 
and at least comparable to zoom-in cosmological simulations \citep[][]{hopkins14,kimm15,agertz15},
GMCs of size $\sim 100\,{\rm pc}$ are resolved only by 20 cells across,
meaning that the internal structure is unlikely to be captured as realistically as 
possible. Although the star formation and outflow rates seem reasonably converged 
(see the Appendix), small-scale simulations  find that the dense pillars of gas 
can self-shield themselves from the radiation, and that star formation continues even 
after a significant fraction of gas in GMCs is over-pressurised and blown 
away by radiation \citep{dale12,dale14,raskutti16}. The width of such dense
filaments from \citet{raskutti16} is $\sim 1\,{\rm pc}$ or so, which is
challenging to achieve in global disk simulations like ours. 
If these internal structures were present in our simulations, 
it might be possible to form more stars per cloud, driving more 
coherent outflows from SNe. However, it is unclear whether there will be enough 
gas left to be entrained near the GMCs or in the disk, as radiation can disperse them 
quite effectively.

Finally, the dynamics of gas in our simulations is governed by 
hydrodynamic interactions and radiation, and any processes associated with 
magnetic fields are neglected. \citet{hennebelle14} shows that magnetic fields 
not only provide additional pressure support that thickens the gaseous disk, 
but also suppress the fragmentation of gas clouds by stabilising the 
Kelvin-Helmholz instability through magnetic tension \citep{ryu00,hennebelle13}.
The resulting star formation rates are reduced by a factor of two. 
Given that star formation
may become more clustered in the presence of magnetic fields and that magnetic 
tension can hold the gas more effectively, it will be interesting to see how the reduction
in star formation interplays with magnetic fields and drives outflows.  
We also note that the cosmic ray pressure has a potential to drive strong outflows. 
As the non-thermal energy cools more slowly ($\gamma=4/3$) than thermal component 
of the ideal gas ($\gamma=5/3$), it is argued that cosmic rays help to build up stable 
vertical pressure gradient and launch warm ($T\sim10^4\,{\rm K}$) 
outflows \citep{uhlig12,booth13,salem14,girichidis16,simpson16,pakmor16}.
However, \citet{booth13} suggests that the mass loading becomes  
on the order of unity in a $10^9\,\msun$ dark matter halo 
when star formation rates become comparable to our simulated galaxies 
($\sim10^{-2}\,\msunyr$). This is certainly non-negligible, but the relative importance 
between different physical processes will be unraveled only by running full-blown 
radiation magneto-hydrodynamic simulations with anisotropic diffusion on galactic scales.
 
\section{Conclusions}
Using a set of radiation-hydrodynamic simulations of a disk galaxy
embedded in a small dark matter halo of mass $10^{10}\,\msun$, 
we investigate the effects of resonantly scattered \Lya\ photons 
on the properties of galaxies  and address the issue whether it is possible to drive 
strong winds by adopting strong radiation feedback. For this purpose, we calculate the 
multiplication factors of \Lya\ photons, 
which boosts the momentum transfer to the gas due to multi-scattering,
on a static, uniform medium
with a variety of metallicities using a MCRT code, {\sc rascas} (Michel-Dansac et al. {\sl in preparation}),
and develop a sub-grid model for the early \Lya\ feedback in the {\sc ramses} code \citep{teyssier02,rosdahl13}.
We also provide fitting functions for the multiplication factors (Eqs.~\ref{eq:mf1}--\ref{eq:mf_dust}) which may be useful to the community.
Our findings can be summarised as follows.
\begin{itemize}
\item In metal-poor systems with low dust content, the multiplication 
factor can be as high as several hundreds (Fig.~\ref{fig:mfactor}).
In this regime, \Lya\ pressure is a more efficient feedback process than 
photo-ionization heating due to LyC photons (Eq.~\ref{eq:ralpha}) or 
direct radiation pressure from UV photons, and comparable in momentum to SNe (Fig.~\ref{fig:mom}). 
\item The momentum input from \Lya\ pressure operates mostly
on cloud scales (Fig.~\ref{fig:pic_ex}), and is able to disrupt star-forming clouds 
before SNe blow out a significant amount of gas.
As a result, the number of SNe exploding in dense environments ($\nH\ga100\,\cmq$) 
is dramatically reduced (Fig.~\ref{fig:snsites}), and the star formation rates 
are suppressed by a factor of two, compared to the RHD run without \Lya\ 
feedback (\texttt{G8R-SN}) (Fig.~\ref{fig:sf}).
\item The most significant effect of strong early feedback is to smooth out 
bursty star formation. This leads to weaker galactic outflows, compared to
the run without \Lya\ feedback (Fig.~\ref{fig:out}). The mass-loading factor 
measured at $|z|=2\, {\rm kpc}$ is $\eta_{\rm out}\sim4$ in the case of 
\texttt{G8R-SN-Lya} run, whereas it is higher ($\eta_{\rm out}\sim8$) in the 
\texttt{G8R-SN} or \texttt{G8SN} runs where star formation is more bursty 
(Fig.~\ref{fig:outdetail}).  \Lya\ feedback alone cannot drive strong, extended 
outflows, as the radiation field drops significantly outside the star-forming clouds.
\item We also test the models in which radiation field strength
is arbitrarily enhanced by a factor of 3 or 10 and a model in which the
star formation efficiency per free-fall time is increased by a factor of 10, 
but the mass-loading factors are quite similar to those 
of the fiducial run (\texttt{G8R-SN-Lya}).
\item The outflows are not mass-conserving when \Lya\ feedback is present.
The mass-loading factors measured at $0.2\,R_{\rm vir}$ 
($=8.2\,{\rm kpc}$) drop to an order of unity in \texttt{G8R-SN-Lya}, whereas 
it remains nearly unchanged in the  \texttt{G8SN} or  \texttt{G8R-SN} runs 
(Fig.~\ref{fig:out_ex}).
\item The inclusion of radiation thickens the disk by providing extra thermal
support, and \Lya\ pressure further increases the scale-height of the disk.
The resulting mid-plane pressure and star formation surface densities 
are good agreement with the models maintaining vertical equilibrium 
(Fig.~\ref{fig:ISM}).
\item While long-lived star clusters are allowed to form in the \texttt{G8R-SN} run 
due to an efficient conversion of gas into stars before the SNe emerge, 
\Lya\ pressure not only reduces the number of star clusters but also 
decreases the typical mass of individual star cluster (Fig.~\ref{fig:clusterimg}).
\item Escape fractions of LyC photons are slightly lower in the run with \Lya\
feedback ($\left<f_{\rm esc}^{\rm LyC}\right>=5.5\%$), compared with the 
SN run  ($\left<f_{\rm esc}^{\rm LyC}\right>=6.9\%$). This is again because 
\Lya\ alone cannot drive strong winds and because SNe become less correlated 
due to less bursty star formation. In contrast, when the radiation field is strong 
enough to dominate over SNe, the escape fractions are increased to 
$\sim$ 12\% (\texttt{G8R-SN-Lya-f3}) or 19\% (\texttt{G8R-SN-Lya-f10}). 
Despite the increase in escape fractions, the total number of escaping LyC photons
remains similar within $\sim 20\%$, as the suppression of star formation balances 
the increase in the escape fraction. This suggests that the reionization history of the universe is likely not 
significantly affected by \Lya\ pressure.
\end{itemize}

\section*{Acknowledgements}
Support for this work was partly provided by the National Research Foundation of Korea 
to the Center for Galaxy Evolution Research (No. 2017R1A5A1070354) and partly by the 
ERC Advanced Grant 320596 ``The Emergence of Structure during the Epoch of Reionization".
JR and JB acknowledge support from the ORAGE project from the Agence Nationale de la 
Recherche under grand ANR-14-CE33-0016-03. HK is supported by Foundation Boustany, 
the Isaac Newton Studentship, and the Cambridge Overseas Trust. 
TG is grateful to the LABEX Lyon Institute of Origins (ANR-10-LABX-0066) of the Université de Lyon for its financial support within the programme 'Investissements d'Avenir' (ANR-11-IDEX-0007) of the French government operated by the National Research Agency (ANR). 
This work used the DiRAC Complexity system, operated by the University of Leicester IT Services, 
which forms part of the STFC DiRAC HPC Facility (www.dirac.ac.uk).   This equipment is funded by 
BIS National E-Infrastructure capital grant ST/K000373/1 and  STFC DiRAC Operations grant ST/K0003259/1. 
DiRAC is part of the National E-Infrastructure.

\bibliographystyle{mnras}
\bibliography{refs}

\section*{Appendix}
\begin{figure}
   \centering
     \includegraphics[width=8.cm]{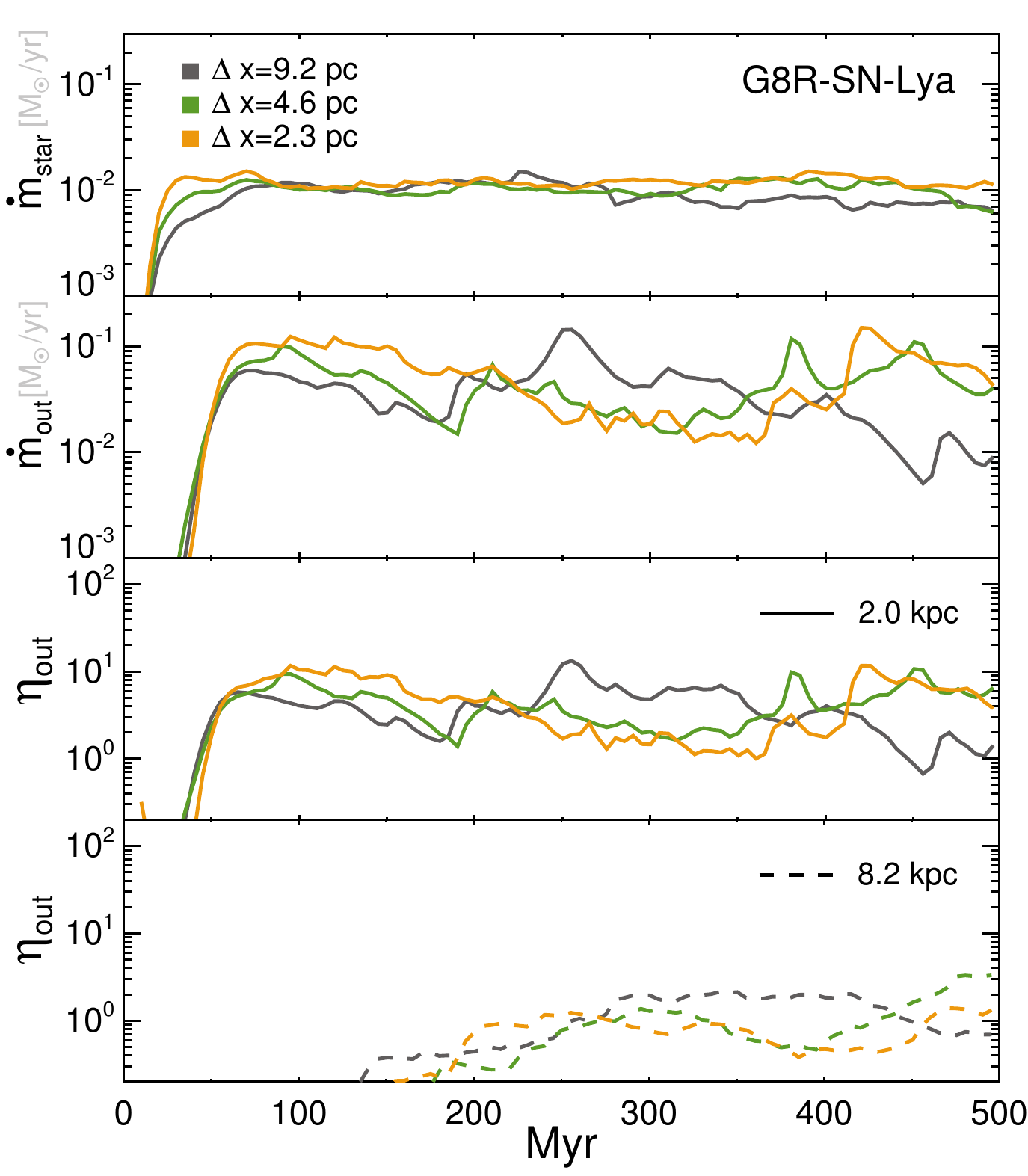}
   \caption{ Resolution test of the runs with \Lya\ pressure. From top to bottom,
   each panel shows the star formation rates in $\msun\,{\rm yr}^{-1}$,
   the outflow rates measured at $|z|=2\,{\rm kpc}$, the mass-loading factors 
   measured at $|z|=2\,{\rm kpc}$ and $|z|=0.2\,R_{\rm vir}\, (8.2\,{\rm kpc})$.
   }
   \label{fig:resolution}
\end{figure}

We note that our resolution (2.3--4.6 pc) may not be high enough to resolve the detailed turbulent structure inside a GMC, 
and the propagation of ionization fronts may be affected by the finite resolution. However, the eventual (D-type) expansion of 
a HII region in dense environments is driven by the over-pressure inside the bubble, and the radial extent to which photoionization heating or \Lya\ pressure can counterbalance the ambient pressure is generally larger than our resolution (Eq. \ref{eq:ralpha}--\ref{eq:rph}). 
Therefore, we expect that the star formation histories of our simulated galaxies should not be affected by resolution. 
To demonstrate this, we perform resolution tests in Figure~\ref{fig:resolution}. Specifically, we run two additional simulations with one more 
 and one fewer levels of refinement (minimum cell widths of 2.3 pc and 9.2 pc, respectively), keeping other parameters
fixed, except for the minimum gas mass to trigger the refinement in the high resolution
run (125 $\msun$ instead of 1000 $\msun$ for the higher resolution run). 
 It can be seen that both the star formation histories 
and outflow rates are reasonably converged. In the early collapsing phase 
($t\la 50\,{\rm Myr}$), the higher resolution runs show systematically higher 
star formation rates, because the gaseous disk fragments more efficiently. 
Accordingly, outflow rates are slightly more enhanced in the higher resolution runs 
during the first $\sim 200\,{\rm Myr}$. However, as soon as the collapsing phase 
ends, the star formation rates averaged over 50 Myr settle around $\sim 0.01\,\msun\,{\rm yr^{-1}}$, and the mass-loading factors become similar between 
these runs, indicating that the simulation results are converged at several 
pc resolutions. We also check that the properties of the ISM are reasonably
similar, although the high resolution run yields slightly thicker disk (73 pc vs 80 pc) 
due to stronger mid plane pressure ($P/K = 10^{4.7}\,{\rm cm^{-3}\, K}$ vs $P/K = 10^{4.9}\,{\rm cm^{-3}\, K}$ ).

\end{document}